\documentclass[12pt,preprint]{aastex}

\slugcomment{}

\shorttitle{}
\shortauthors{Osten et al.} 

\begin{document}
\title{DRAFTS: A Deep, Rapid Archival Flare Transient Search in the Galactic Bulge} 
\shorttitle{Optical Flares in Cool Bulge Stars} 
 
\author{Rachel A. Osten}
\affil{Space Telescope Science Institute, 3700 San Martin Drive, Baltimore, MD 21218}
\email{osten@stsci.edu}
\author{Adam Kowalski}
\affil{Astronomy Department, Box 351580, University of Washington, Seattle, WA 98195}
\author{Kailash Sahu}
\affil{Space Telescope Science Institute, 3700 San Martin Drive, Baltimore, MD 21218}
\author{Suzanne L. Hawley}
\affil{Astronomy Department,  Box 351580, University of Washington, Seattle, WA 98195}

\begin{abstract}
We utilize the Sagittarius Window Eclipsing Extrasolar Planet Search (SWEEPS)
HST/ACS
dataset for a Deep Rapid Archival Flare Transient Search (DRAFTS)
to constrain the flare rate toward the older stellar population
in the Galactic bulge.  During 7 days
of monitoring 229,293 stars brighter than V=29.5, we find evidence
for flaring activity in 105 stars between V=20 and V=28.
We divided the sample into non-variable stars
and variable stars whose light curves contain large-scale variability.
The flare rate on variable
stars is $\sim$ 700 times that of non-variable stars, with a significant
correlation between the amount of underlying stellar variability and peak flare
amplitude.
The flare energy loss rates
are generally higher than those of nearby well-studied single dMe flare stars.
The distribution of proper motions is consistent with the flaring stars being 
at the distance and age of the Galactic bulge.
If they are single dwarfs, they span a range of $\approx$ 1.0 $-$ 0.25M$_{\odot}$.
A majority of the flaring stars exhibit periodic photometric modulations with
P $<$3d.  If these are 
tidally locked magnetically active binary systems, their fraction in the bulge is enhanced by a factor of $\sim$20 compared to
the local value.
These stars may be useful for placing constraints on the angular momentum evolution
of cool close binary stars.
Our results expand the type of stars studied for flares in the optical band, and
suggest that future sensitive optical time-domain studies will have to contend with a larger
sample of flaring stars than the M dwarf flare stars usually considered.
\end{abstract}


\section{Introduction}
Stellar flares are the most dramatic example of variability seen in stars with outer
convective envelopes during the bulk of their lifetimes.  
The study of stellar flares can inform several topics: understanding
the observable manifestations of magnetic reconnection in stellar
atmospheres; constraining the expected stellar flare rate  
in optical transient searches for rare phenomena;
and exploring the importance of stellar flares on exoplanet
environment and habitability.
The study of solar white-light flares has revealed the importance of 
the UV and optical regions in the energetics of flares; although 
flare variations at other wavelength regions such as X-rays are larger,
white light continuum emission dominates the total radiated energy, by factors
of roughly 100 with respect to X-ray energies \citep{emslie2005,kretzschmar2011}.

The general consensus
appears to be that stellar flares are produced by the same basic physical
processes occurring in solar flares.
This is despite the fact that flaring stars
can be significantly different 
from the Sun: F dwarfs \citep{mm2000},
G and K giants \citep{ayres1999,ayres2001,testa2007},
tidally locked RS CVn binary systems \citep{osten2004}, dMe 
flare stars \citep{osten2005}, flares on hyperactive young Suns containing 
star-disk interactions \citep{favata2005}, and very low mass dwarfs near the 
substellar limit \citep{stelzer2006}.
At optical wavelengths, the selection for significant brightness increase 
in broad band filters means that M dwarfs are the primary observed flaring stars,
due to the intrinsic red stellar spectrum compared with the 
blue flare spectrum \citep[flare continua are characterized roughly by 
a blackbody of T$\sim$10$^{4}$K compared with stellar T$_{\rm eff} \lesssim$3500 K;][]{hawley2003}.
dMe stars are known to undergo dramatic optical variations (peak enhancements in the U band
can be several magnitudes over timescales of minutes),
with mean flare energies in the U-band of about 10$^{32}$ erg ---
comparable to the largest solar flares.
The flare rates for the most active dMe flare stars are 
0.2--6 flares/star/hr, 10$^{4}$--10$^{5}$ times that for the Sun
\citep{lacy1976}.
Optical flare timescales on M dwarf flare stars can range from less than a minute to several tens of 
minutes
\citep{hawley2003}
 with
mean flare duration independent of flare amplitude 
\citep{kunkel1973}.

Due to their dependence on the presence and dynamics of
interacting magnetic fields, stellar flares are a diagnostic
of transient magnetic activity, in addition to more commonly used 
persistent indicators such as chromospheric/coronal emissions,
starspots and rotation.  In situations where persistent magnetic activity
is lacking, transient magnetic
activity provides a complementary diagnostic of the presence of large-scale magnetic fields.
Magnetic activity is a function of convection zone depth
and rotation rate, as quantified by the Rossby number 
\citep{noyes1984}, for
stars that are not fully convective.
This relates ultimately to the dynamo mechanism
producing the large-scale magnetic fields whose stresses and
plasma interactions cause the mechanical heating and reconnection
observed as different magnetic activity phenomena.
In single solar-like stars, magnetic activity is a function of age,
with young stars exhibiting enhanced magnetic activity, and
a consequent fall-off with increasing age accompanying spin-down
\citep{skumanich1972,soderblom1982}.
Likewise, the stellar flare rate appears to decline as a function of age.
Results for low mass stars have suggested that 
activity timescales for fully convective stars may be as large as 10 Gyr
\citep{delfosse1998,west2008}.
While there have been several recent studies concentrating on 
the flaring rate in young stars \citep{wolk2005,stelzer2000},
the decline in activity at Gyr ages means that relatively short
exposures of single stars will generally not reveal any flaring activity,
and the flare rate in older stars has not been studied systematically.
Statistical studies of stellar magnetic activity have found that
the flare rate decreases with height above the Galactic plane, 
as dynamical effects cause stars to move farther away from the plane as they age
\citep{welsh2007,kowalski2009,hilton2010,keplerref}.

Stellar flares can also occur in binary systems.
For tidally locked binaries, the tidal interaction
forces the stars to rotate rapidly, in synchrony with the orbital period. 
Several classes of binary systems 
with at least one cool member are tidally locked (notably the RS CVn, BY Dra, and W UMa type binaries)
and display enhanced chromospheric and coronal activity, and the RS CVn and BY Dra cases 
commonly 
demonstrate large individual flaring events \citep{guinan1993}.  Because the tidal interaction causes
the stars to rotate much more rapidly than if they were single, activity (including flaring)
can be maintained in much older systems than for single stars.
Algol binary systems, in which some mass transfer is taking place, can also display magnetic activity;
as the spectral types of the cool secondary are similar to those seen in the RS CVn binaries,
it is understood that the magnetically active secondary star is the site of such activity \citep{guinan1993}.

Recent results from searches for deep optical fast ($<$1 hr) transients have indicated the prevalence
 of stellar flaring
and bolstered support for a better quantification of the flare rate.  
The three transients identified in \citet{becker2004} turned out to be Galactic flare stars;
\citet{kulkarni2006} described this population as a foreground ``fog'' potentially obscuring 
extragalactic
fast optical transients (timescales of $\approx$1000 s).
\citet{fresneau2001} reported results from an astrographic plate survey at
low galactic latitude, covering a large field of view (520 degrees$^2$) at B magnitudes between 10 and
14; 8\% of stars
showed flare events greater than 0.4 mag over 20-30 minute timescales.
\citet{ramsay2005} performed sensitive but shallow temporal coverage observations at 
intermediate galactic latitude, reaching depths of V$\sim$22.5 on $\sim$ minute timescales.
With only 2 hours per field, they found 45 variable targets with 2 flare-like objects. 
From V $>$ 16.5 mag to 22.5, $\approx$ 0.2\% of
stellar sources were variable.  
This is especially problematic for sensitive transient searches like PanSTARRS and LSST, the
latter possibly having as many as 50 stellar flares in each $u$ band exposure \citep{hilton2010cs}.


On normal stars, we expect temporal variations in the stellar light to come from 
surface activity, as well as from
eclipses by stellar or substellar companions
and variations due to seismic waves from the stellar interior.
Flares are stochastically occurring, and tend to have fast timescales, although large long-duration 
flares can last for hours.  On the Sun, white-light flares have typical durations of only a few minutes,
with small enhancements \citep[$>$10ppm;][]{kretzschmar2011}.
A second source of temporal variations are due to the rotation of surface features
on and off the
stellar disk.  Concentrations of surface magnetic field 
can produce intensity variations both lighter (such as faculae) and darker (such as spots)
than the surrounding photospheric material:
these features can be long-lived,
producing regular photometric variations over multiple rotation periods if the
size and location do not change appreciably during its lifetime \citep{hussain2002}.
Irregular periodic variations could be due to growth and evolution of these surface features.  
Coupled with differential
rotation of the stellar surface, these effects add complexity
to the periodic variations beyond a simple sinusoidal dependence.

Because the flare rate is expected to be low on older stars, a systematic search for
flares in an older stellar population needs a large number of stars,
and involve a relatively long stare coupled with fast cadence to detect and resolve the flaring
emission from any other variability.
\citet{sahu2006a} announced the discovery of 16 candidate extrasolar planets
using 
nearly continuous 7 day Hubble Space Telescope/Advanced Camera for Surveys (HST/ACS) monitoring of one field 
of 229,701 stars in two filters.
The motivation for the initial study was to look for transiting extrasolar planets 
with lower masses and a larger range of metallicities in a part of the Galaxy that
had not been explored previously.
The color-magnitude diagram of the stars shows that the field is dominated by old stars
associated with the Galactic bulge, as well as younger objects from the foreground disk population.
The bulge population has an age estimated from isochrone fitting of $\sim$ 10 Gyr.
Due to the characteristics of the stellar target sample and observing strategy, 
this dataset is ideal for the purposes of studying the incidence
of stellar flares in an older stellar population and comparing to the decline of magnetic activity
seen in younger stellar populations, and thus forms the basis of the archival project
described in this paper.  The paper is laid out as follows: \S 2 summarizes the data reduction,
\S 3 describes the analysis of the light curves, \S 4 describes the results, \S 5 discusses the 
implications, and \S 6 summarizes.

\section{Data Reduction}
The data used in this paper originally came from the SWEEPS project, described in 
\citet{sahu2006a}.
The SWEEPS field (202''x202'') was imaged continuously for 7 days in 2004 February.  
The observations consist
of 254 exposures in V (F606W) and 265 in I (F814W), each with an exposure time of 339 seconds,
and typical spacing between individual observations $\sim$8 minutes. 
The time-series photometry was obtained through the use of difference imaging
and is described in \citet{sahu2006a}.  For each star, a time series of relative intensities
was constructed.
Because the contrast of flares to the surrounding
photosphere is highest at shorter wavelengths, we concentrate primarily on the $V$ band time
series in this paper.
The magnitude and color of each star
were used to estimate stellar parameters such as mass and temperature.  
The time series data are already calibrated and provide the starting point to examine evidence for 
variability
and flares. 

\section{Analysis}
The total number of stars in the SWEEPS sample brighter than V=29.5 is 229,293.
After perusal of the light curves, a subset was retained where more than 70\% of the light curve
points have error bars no larger than three times the standard deviation.
This reduced the number of stars to 222,657.  
A further cut was made on stars fainter than V=20 to select dwarf stars, reducing the
number to 216,136.
This serves as the starting point for our flare transient search, which we describe in more detail
below.  

\subsection{Sample Division and Detrending of Variable Stars}
\begin{figure}
\includegraphics{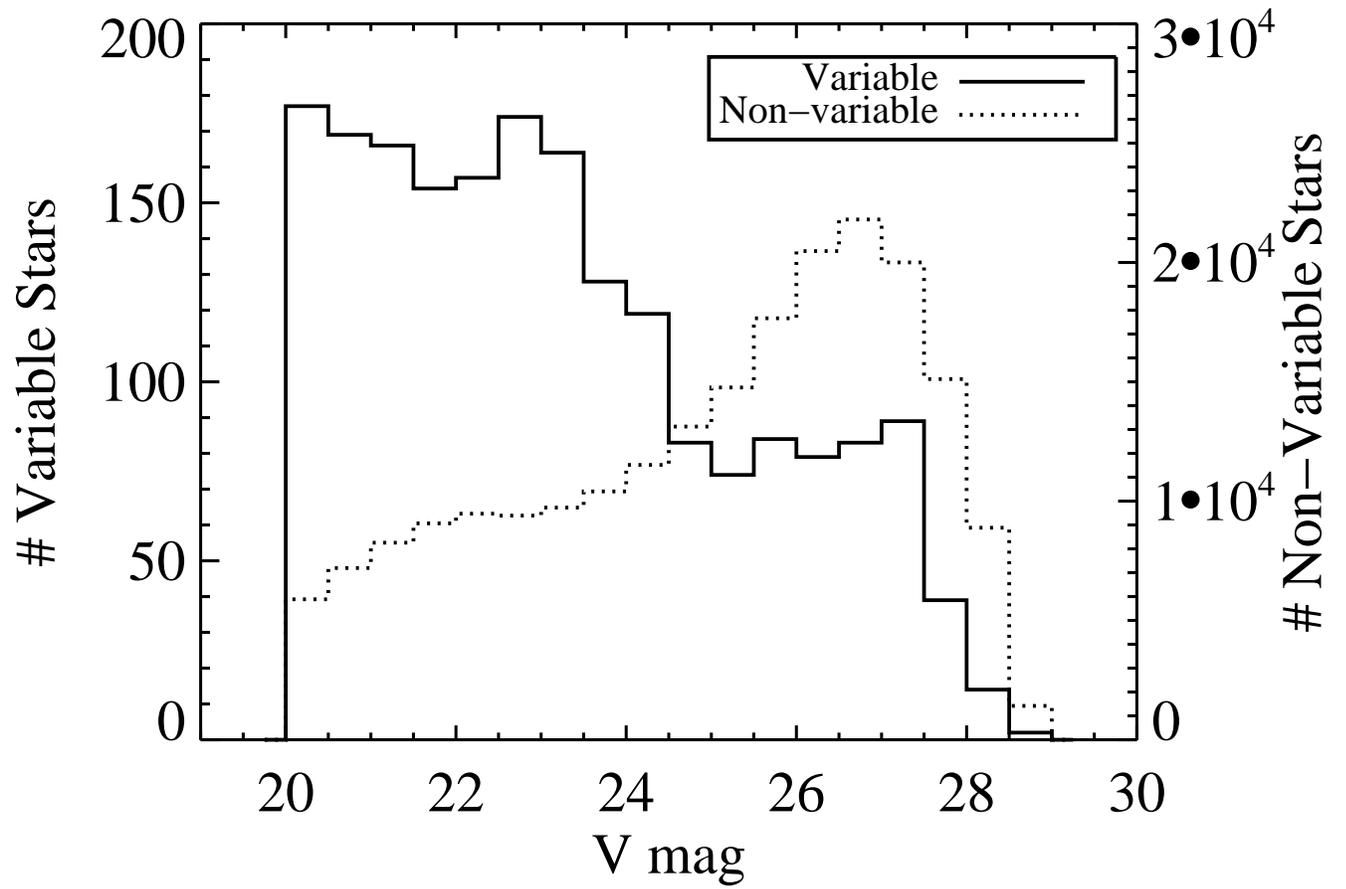}
\caption{Plot of distribution of stars as a function of V$_{\rm F606W}$ magnitude,
for stars deemed variable and non-variable, respectively.  See
\S3.1 for discussion of sample division.  \label{fig:varnonvar}
}
\end{figure}

The data format is a record of the relative flux of each star, where V$_{\rm rel,i}$
is  the relative flux of the $i$th time bin, defined as \\
\begin{equation}
V_{\rm rel,i} = \left( \Delta F/F \right)_{i} = \left( \frac{F_{i}-\overline{F}}{\overline{F}} \right)_{V}
\end{equation}
where $F_{i}$ is the flux measured in time bin $i$, and
$\overline{F}$ is the average flux computed from the entire time series
for a given filter (here, the V606W filter).
I$_{\rm rel,i}$ is defined similarly for the F814W filter.
The quantity $V_{\rm rel,i}$ is also expressed as $\Delta F/F_{i}$ in some figures in the paper.
In order to characterize the light curves to search for transient variability, we
determined a number of statistics.  The first cut utilized the reduced chi-squared statistic
of the relative flux summed over the time series.
For these purposes $\chi^{2}_{\nu}$ is defined as \\
\begin{equation}
\chi^{2}_{\nu} = \frac{\sum_{i} V_{\rm rel,i}^{2}/\sigma_{i}^{2}} {(N_{\rm lc}-1)}
\end{equation}
where $V_{\rm rel,i}$ is defined above, $\sigma_{i}$ is the uncertainty
associated with the measurement, and $N_{\rm lc}$ is the number of points in the light curve. 
Stars with a reduced chi-squared less than 1.5 were termed non-variable, and
those with $\chi^{2}_{\nu}>$1.5 variable. The numbers of non-variable and
variable stars using this definition are 214,181 and 1955, respectively.
Figure~\ref{fig:varnonvar} displays the distribution of these two classes as a function
of stellar V magnitude.

\begin{figure}
\includegraphics[angle=90,scale=0.7]{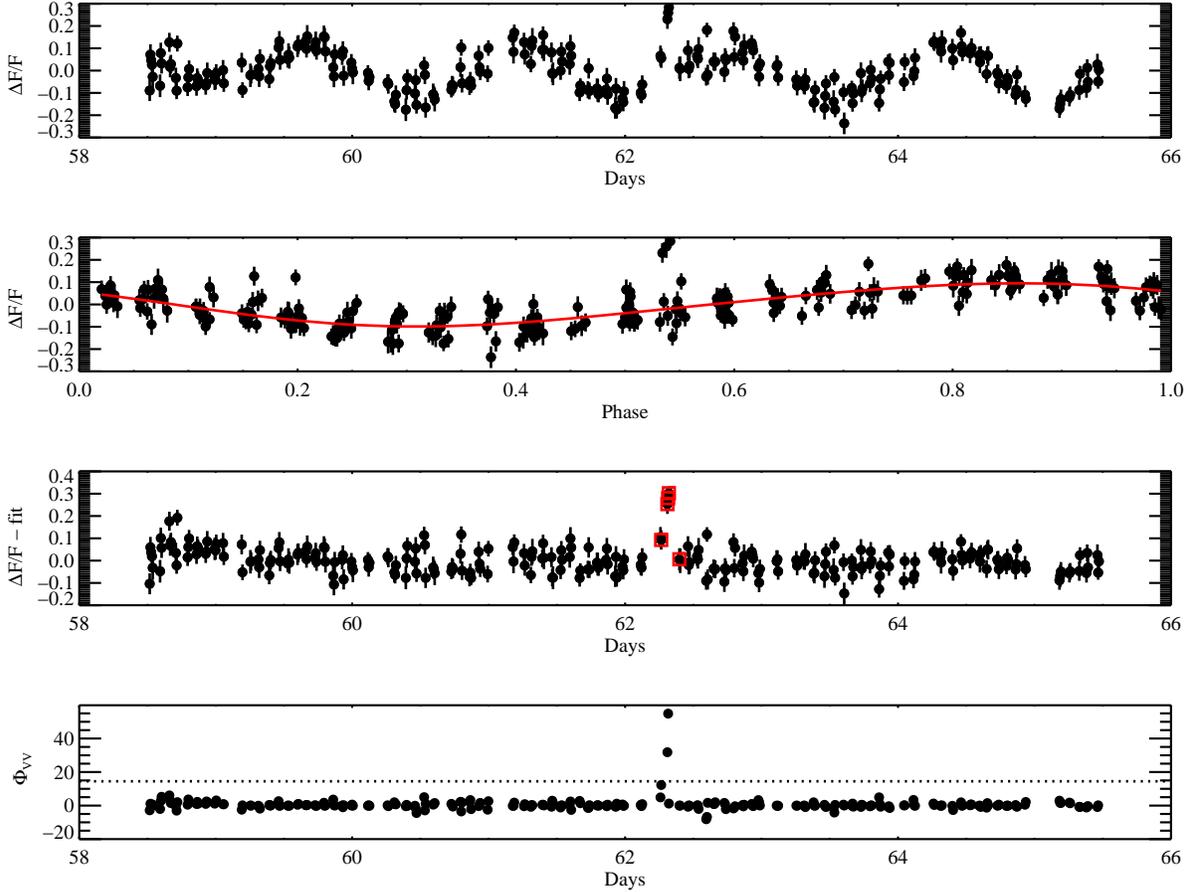}
\caption{Example of a flare identified on a star with a variable light curve exhibiting 
regular periodic variations.  Top panel shows
relative fluxes versus time, and second panel from top shows the same data points,
but folded over the best-fit period; the fit is shown in red. The third panel from the
top shows the data after subtraction of the fit; red squares indicate time bins identified
as a flare event.  The bottom panel shows the $\phi_{VV}$ values for the detrended light curve.
Only points which lie above the threshold value of $\phi_{VV}$=14.5 (dotted line in the bottom panel), have a value more than 2.5
times the standard deviation, and exhibit a positive increase are identified as flare peaks.  The flare
event (red squares in the third panel from the top) 
is constructed from data points adjacent to the flare peaks which have relative fluxes 
greater than zero.  See \S 3.1 and \S 3.2 for more discussion.
\label{fig:actperflare}}
\end{figure}

\begin{figure}
\includegraphics[angle=90,scale=0.7]{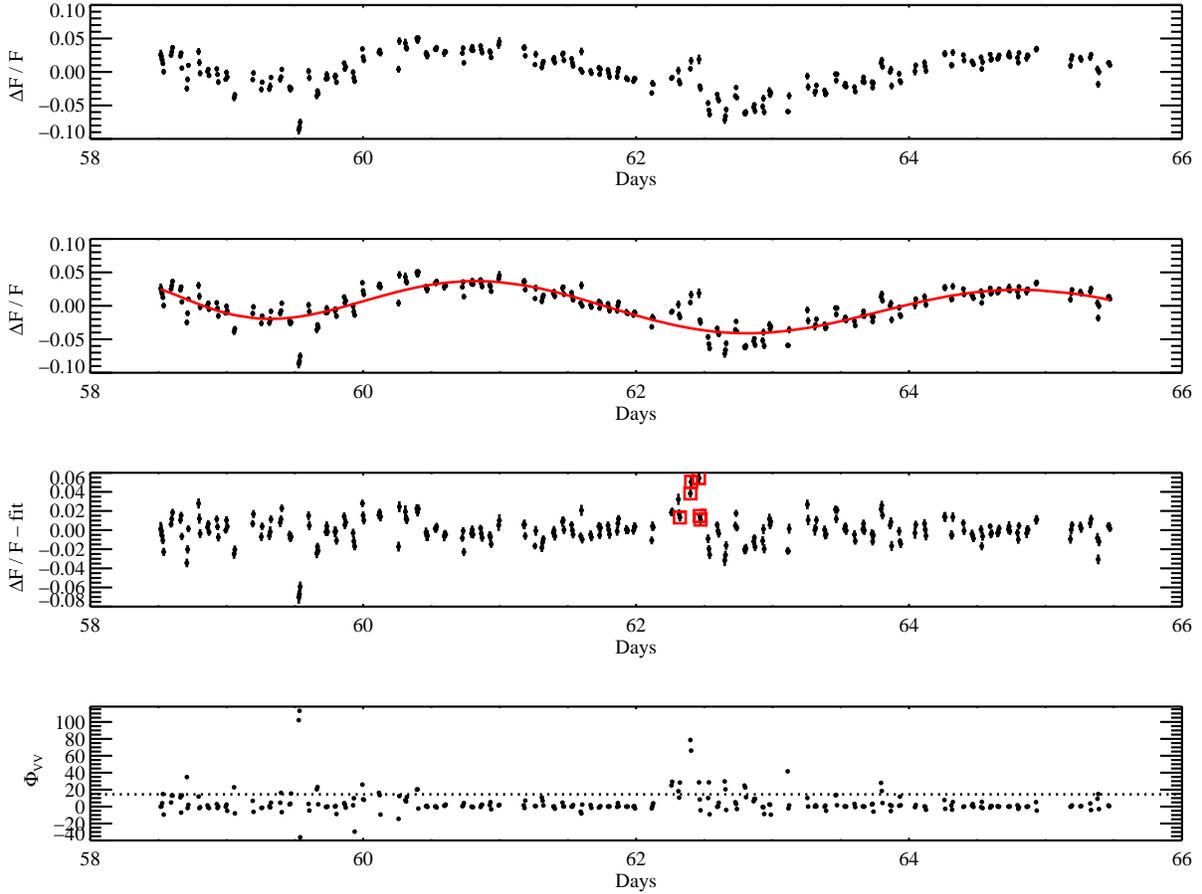}
\caption{ Example of a flare identified on a star with irregular variations.
Panels are as in Figure~\ref{fig:actperflare}.
The irregularly variable stars may have varying amplitudes, such that when phase-folded over
the dominant period these amplitudes don't line up, or a changing slope to the rise of each light curve
portion.   In this example, the rise in amplitude from $t=60-61$ is different than from $t=63-65$, and
the peak amplitudes are different.
The fitting methods described in \S3.1 are designed to detrend the light curve so that 
smaller scale temporal variations  (flares) can be identified and studied.
See \S3.1 and \S3.2 for more discussion.
\label{fig:actnonperflare}}
\end{figure}

The variable star sample had underlying
large-scale variations (large amplitude and long duration relative to single epochs), 
which needed to be removed in order to see any evidence of short-term
flare variations. 
The following two approaches were applied to all
light curves in the variable star sample, and the best fit was retained.
In the first approach, periodic variations were identified using a Lomb normalized periodogram, 
and the light curve was folded to
the dominant period. 
The light curve was then fit by a sum of up to 4 sine terms:\\
\begin{equation}
V_{\rm rel,i}=\sum_{n=1}^{max1} a_{n} \sin (2\pi n (\phi_{i}+\rho_{n})) \label{eq:1}
\end{equation}
where V$_{\rm rel,i}$ is the relative flux at epoch $i$, $\phi_{i}$ is the phase of epoch $i$ when
folded over the dominant period, $\rho_{n}$ is a phase offset, and $a_{n}$ is a normalization.
The number of sine terms $max1$ was set to 2 or 4, following \citet{pojmanski2002}.
An F-test determined whether the additional two terms produced
a statistically significant decrease in the chi-squared statistic.
In the second approach, the light curves were fit by a sum
of sines with up to 8 terms, using the following prescription:\\
\begin{equation}
V_{\rm rel,i}=\sum_{n=1}^{max2} a_{n} \sin (2\pi n (\tilde{t}_{i}+\rho_{n})) \label{eq:2}
\end{equation}
with \~{t} a normalized time ($\tilde{t}_{i}=t_{i}/(t_{\rm max} - t_{\rm min})$)where
$t_{\rm max}$ and $t_{\rm min}$ are the last and first times in the time series, respectively,
and
$max2$ was $2$, $4$, $6$, or $8$, giving up to
16 free parameters.
The value of $max2$ was taken to be larger than $max1$ to allow more flexibility
with the irregularly variable stars.
We performed fits to optimize the parameters.
A decision as to which prescription (described by equation~\ref{eq:1} or ~\ref{eq:2})
 was better
was made by comparing $\chi^{2}$ and number of degrees of freedom $\nu$ through an F-test.
The fit was then subtracted
from the original time series 
to detrend the light curve.
We termed light curves fit better by the first approach regularly variable,
and those described better by the second approach irregularly variable.
A total of 1837 out of the 1955 variable objects had successful detrending. 
Of the 1955 variable objects, 1443 were regularly variable and 512 irregularly variable;
the number changed to 1335 and 502, respectively, for the variable objects with 
successful detrending.
The majority of the regularly variable objects discarded due to poor detrending had
very short ($<$0.4 d) periods. 
The first two panels of Figure~\ref{fig:actperflare} illustrate the detrending process
for a regularly variable star, while the first two panels of Figure~\ref{fig:actnonperflare}
illustrate the process for an irregularly variable star.

\subsection{Flare Selection}
We tested a couple of different statistics in developing our algorithm to find flares.
Initially we used an $n\sigma$ cut, filtering on individual points which were outliers
from $V_{\rm rel}=0$ at the level of $n\sigma$. With $n\gtrsim$5 and $\sigma$ the standard deviation
of a normally distributed dataset, Gaussian statistics indicates a low 
probability of such events happening, and this method can be used to identify
outliers which may be flares.
 However, there are potentially
a large number of cosmic ray hits which may show up as outliers using this method, and while it is theoretically
possible to separate a true flare which is occurs in a single time bin from a cosmic ray by using the PSF shape
of that exposure for that star,
this method returns an unwieldy number of potential flares.
We decided instead to employ a different set of criteria based on 
adjacent time bins.  
We define a statistic $\phi_{VV}$ 
based on 
the work of \citet{welch1993} and \citet{stetson1996}, 
as follows:\\
\begin{equation}
\phi_{VV} \;\;\; =\left( \frac{V_{\rm rel}}{\sigma} \right)_{i} \times
\left( \frac{V_{\rm rel}}{\sigma} \right)_{j}
\end{equation}
where $V_{\rm rel,i}$ is the relative flux in epoch $i$ (after detrending, if necessary;
see \S3.1),
and $\sigma_{i}$ is the error on the measurement in epoch $i$.
This computation is done for each subsequent pair of temporal data, so that $j=i+1$.
Whereas \citet{welch1993} and \citet{stetson1996} developed this index to investigate
variability in multi-band and multi-epoch photometry, 
the $\phi_{VV}$ statistic here is applied to subsequent epoch pairs
in the same band
in our database. A similar technique
was used in the analysis of M dwarf flaring in the Sloan Digital Sky Survey Stripe 82 dataset
discussed by \citet{kowalski2009}.
In order to limit the number of false detections, we applied the false discovery rate
analysis of \citet{miller2001} to select flare candidate epochs by using a $\phi_{VV}$ threshold.

\begin{figure}
\begin{center}
\includegraphics[scale=0.5,angle=90]{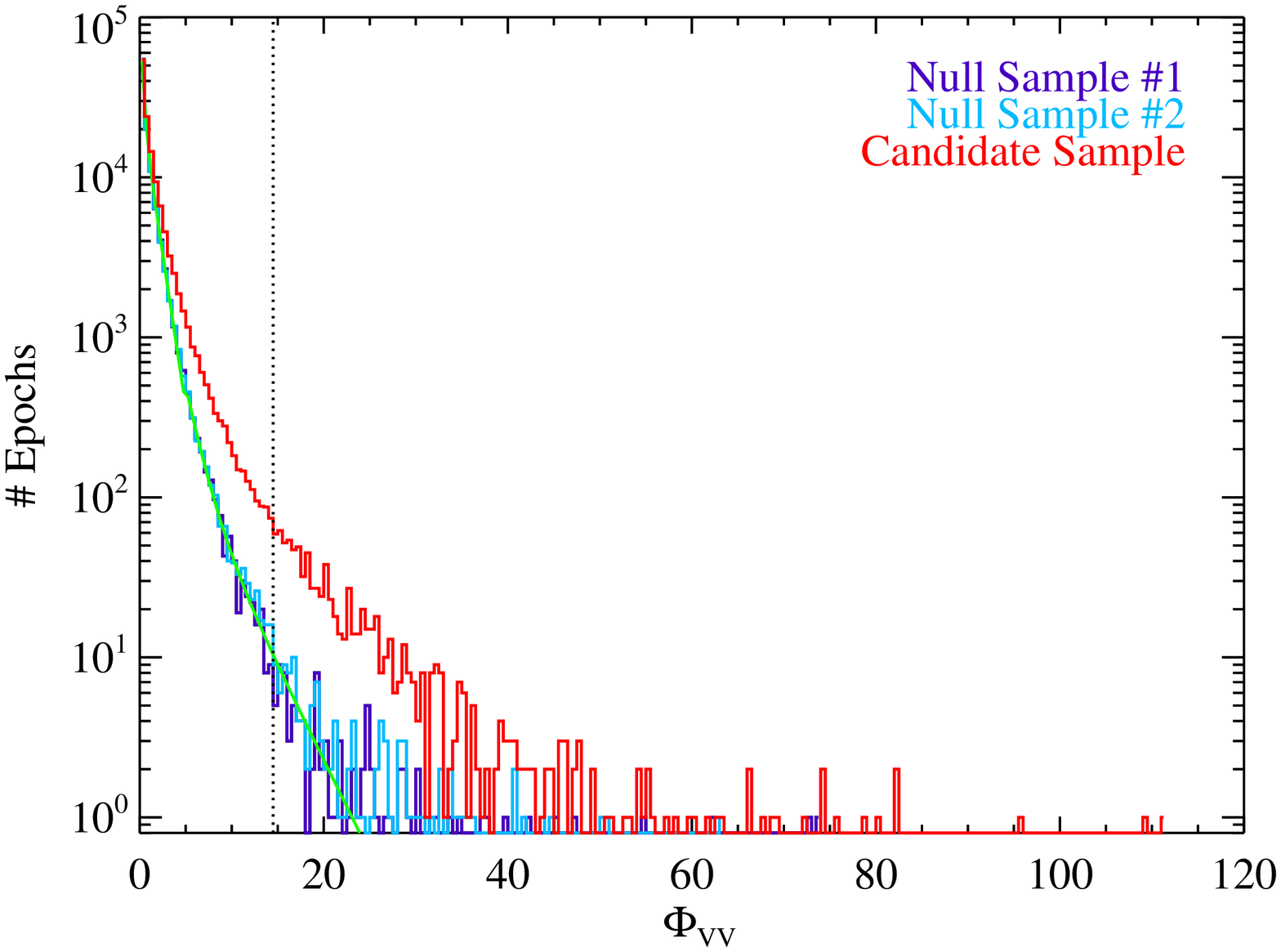}
\includegraphics[scale=0.5,angle=90]{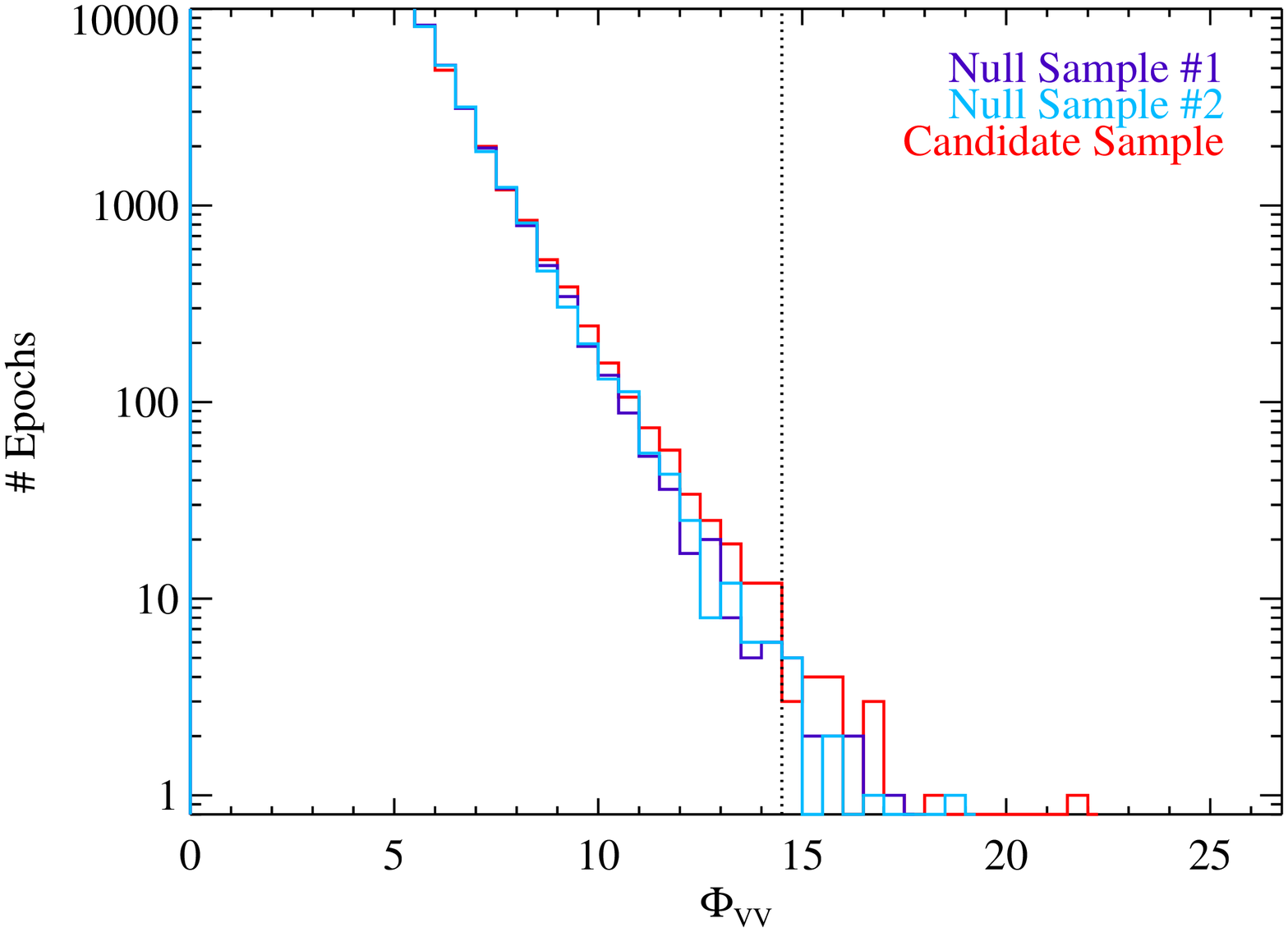}
\end{center}
\caption{False discovery analysis of the variable sample
{\it (top)} and non-variable sample {\it (bottom)}.  
The null distributions (null samples \#1 and \#2) are epoch pairs with
negative and positive flux changes (leading to negative $\phi_{VV}$ values -- their absolute
value is plotted here); the
two null samples differentiate the ordering of the flux changes. 
The green curve in the top panel shows the sum of two Gaussians fitted to the null sample.
The vertical dotted line indicates the cut-off value used to ensure a false discovery rate
of 10\%.
See \S 3.2 for more discussion.
\label{fig:fdr}}
\end{figure}

False-discovery rate analysis \citep[][]{miller2001,hopkins2002}
allows us to set a critical
threshold value of $\Phi_{VV}$ to select flare epochs while ensuring 
a given percentage of false-positives.  
The FDR technique has been recently employed for flare rate
studies in \citet{kowalski2009} and \citet{hilton2010}, as well as other
astrophysical applications.
We begin by dividing the variable detrended (excluding those stars with
poor detrending) star sample distribution of $\Phi_{VV}$ pairs 
(464,761 epoch pairs) into a null distribution (non-flaring
epochs produced from spurious deviations in the data) and a candidate 
distribution (a mix of real flares and spurious non-flares).  For the purposes of 
this calculation we exclude from consideration epoch pairs in which $\Phi_{VV}>0$
but the values of $V_{\rm rel,i,j}<$0.
Specifically, the null distribution is composed of the epoch-pairs consisting of a positive
flux change in one epoch and a negative flux change in the
other epoch (\emph{i.e.}, negative values of $\Phi_{VV}$; 207,381 $\Phi_{VV}$ pairs).
The null distributions for positive then negative flux changes was tallied separately
from the distribution for negative then positive flux changes, to gauge similarity.
The candidate distribution consists of the epoch pairs 
with a positive flux enhancement in both epochs (positive values of $\Phi_{VV}$ and V$_{\rm rel,i,j}>0$; 130,117 $\Phi_{VV}$
pairs).  The sum of these two numbers is smaller than the total number of epoch pairs due to exclusion of eclipses,
which have a positive $\Phi_{VV}$ but are negative in both epochs.
In Figure~\ref{fig:fdr}, we plot the absolute value of the
nulls compared to the candidate distribution.  The candidate distribution contains a significant tail
that is not present in the nulls.  Above our FDR-set
threshold (see below), the number of spurious epochs divided by the number of total
epochs (spurious epochs and real flares) is less than or
equal to our pre-determined false-discovery rate, given by the
variable $\alpha$.  The ratio of spurious epochs to total epochs using the null distribution
comprised of negative then positive flux changes was 0.14, and for positive then negative flux
changes it was 0.10.

For $\alpha$, we choose 10\%, based on experience with similar datasets in \citet{kowalski2009}.  
Successful FDR analysis requires a good
understanding (model) of the null.  We fit a double Gaussian
to model the absolute value of the null distribution for the variable star sample (one Gaussian for $\Phi_{VV}<$5.38, one for
$\Phi_{VV}>$5.38) for the $\Phi_{VV}$ range from 0 to 111 (the maximum value of
$\Phi_{VV}$ in the candidate distribution).  This was done to approximate the tail of the
distribution which exhibited an apparent break near $\phi_{VV}$=5.38. 
From the double Gaussian,
we calculate a p-value for each value of $\Phi_{VV}$ in the candidate
distribution, where the p-value for a $\Phi_{VV}$ bin represents the probability that a null
epoch-pair has that value of $\Phi_{VV}$ or greater.  Then, given our pre-determined $\alpha$,
the FDR IDL prescription in \citet[][; Appendix B]{miller2001} sets the critical
threshold.  
For $\alpha=10$\%, the critical threshold is $\sim$14.5.
This limits the candidate distribution from 130,117 epoch pairs with $\phi_{VV}>$0 and
V$_{\rm rel,i,j}>$0 to 920
epoch pairs.  
We additionally apply the criterion that the flux divided by
the standard deviation in the detrended light curve be $>$ 2.5 in order
to guard against light curves with marginal detrending.
This limits our variable-star flare sample to 229 epoch pairs for by-eye inspection.
We apply the same $\Phi_{VV}$ and standard deviation cuts to the
non-variable star sample.  The same distribution of null vs candidate
epochs for the non-variable star sample is shown in Figure~\ref{fig:fdr}.

\begin{figure}
\includegraphics[bb=300 50 550 730,clip,angle=90,scale=0.7]{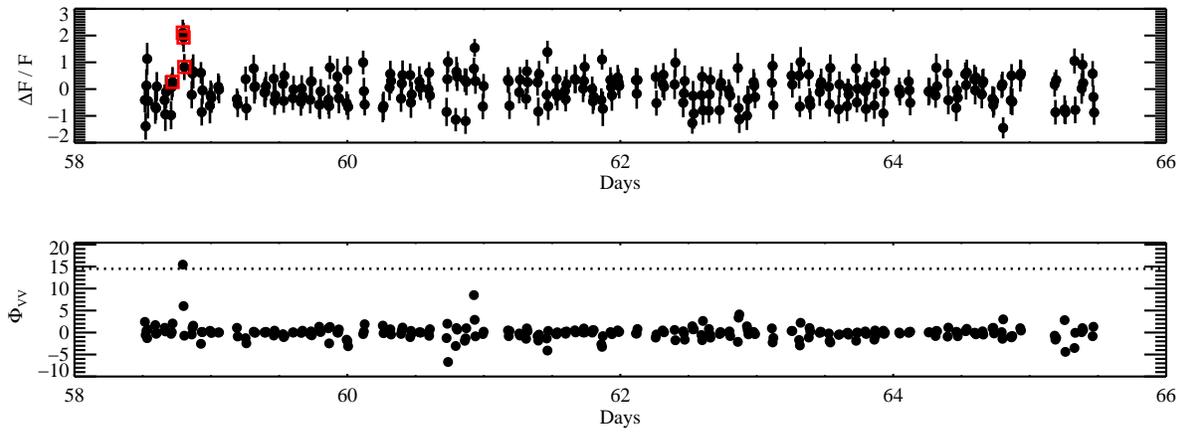}
\caption{ Light curve from a non-variable star, showing flare peak identification and
points associated with a flare event.  The top panel shows the relative flux variations
as a function of time; the bottom panel shows the associated value of $\phi_{VV}$
calculated for each time bin.  
The dotted line in the bottom panel indicates the threshold above which 
peak flare events are identified.  All points on either side of this identified flare
peak which have relative
fluxes greater than zero are included in the flare event.
The red squares in the top panel indicate points identified
as a flare event.  
\label{fig:inactflare}}
\end{figure}

The bottom two panels of Figures~\ref{fig:actperflare} and ~\ref{fig:actnonperflare} 
demonstrate the flare identification technique on variable stars, while Figure~\ref{fig:inactflare}
demonstrates the process on a non-variable star. 
%
Out of 214,181 stars in the non-variable sample, 17 epoch pairs survived using this rigorous
filtering.  In the variable star sample of 1837 stars, 229 epoch pairs were retained.
The next step was to visually inspect all candidate flare epoch pairs.
The above steps were done using $V_{\rm rel}$, and we examined the variations in 
$I_{\rm rel}$ at this stage as well.  We eliminated suspect flares, examples of which
include flares occurring near times of particularly poor detrending (such as at the peak
of a longer-lasting rise in intensity)
or on light curves with
very large scatter in general.
This further reduced the number of
candidate flare epochs.  At this stage it became apparent that systematically the first,
42nd and 43rd epochs were showing up as flares, and we removed these as well.  
This leaves
us with 16 epoch pairs on 16 stars in the non-variable light curves, and 106
flare events (128 flare detections) on the variable light curve sample
(from 89 stars),
for a total of 122 flares from 105 stars.  Of the 89 variable stars exhibiting
flares, 63 had regular periodic variations and 26 were irregularly variable.
Table~\ref{tbl:nbrs} gives a tabular breakdown of the flare identification process
described above.


\section{Results}
\subsection{The Nature of the Flares}
\begin{figure}
\includegraphics[scale=0.7,angle=90]{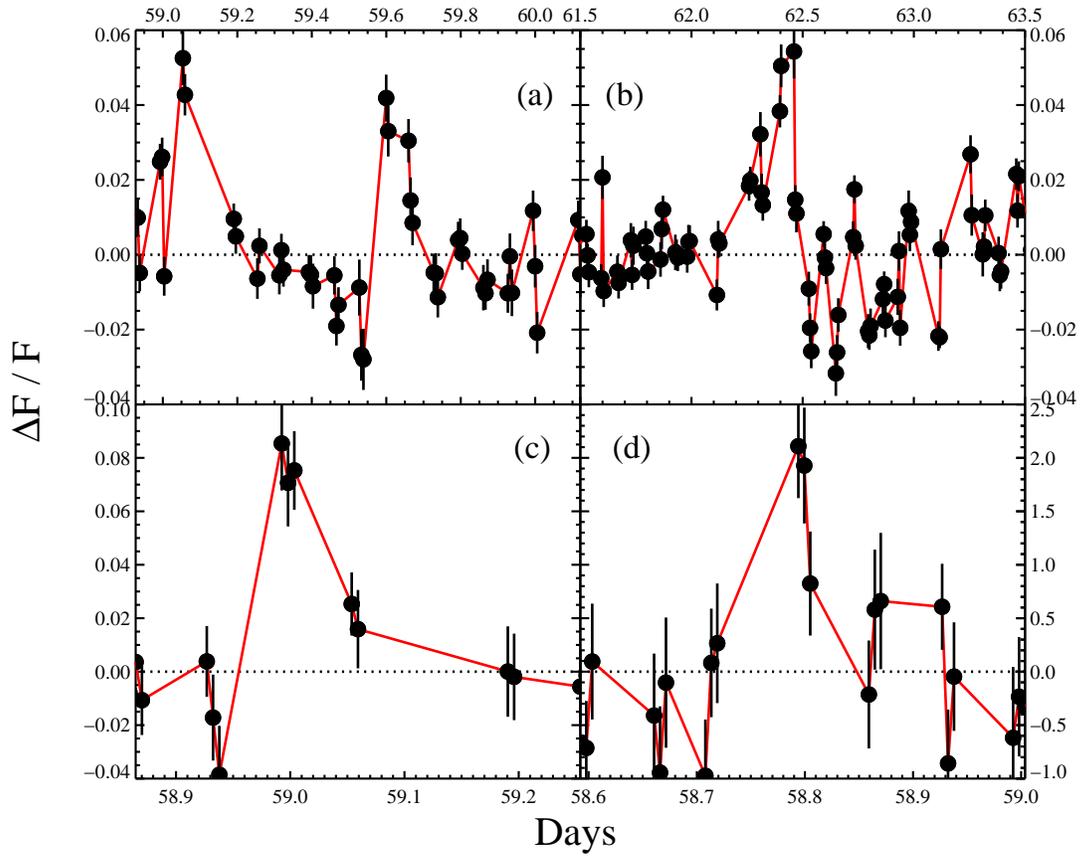}
\caption{Close-up of five sample flares.  
Panels (a), (b), and (d) show flares which occurred on stars exhibiting
underlying variability, and panel (c) depicts a flare seen on a non-variable 
star. Light curves on variable stars have been detrended,
as described in \S 3.1.
Panel (a) shows two flares on the same star while panel (b) shows 
a complex flare event. 
\label{fig:sampleflares}
}
\end{figure}

Once the candidate flares were identified, we determined their properties.
The flare selection described in the preceding section is primarily sensitive
to the flare peaks, since that is where the signal is highest.  In order
to estimate flare properties such as total radiated energy
and average flare luminosity, we considered points with $V_{\rm rel}>0$ 
on either side of the 
adjacent pair identified by the $\phi_{VV}$ statistic 
to define the flare duration.  By the definition of the $\phi_{VV}$ statistic
(see \S 3.2), each flare has a minimum of two points in it.
Figure~\ref{fig:sampleflares} illustrates some example flares.

We summed the intensities in the exposures identified as belonging to the flare
and used the stellar luminosity to determine the flare energy in the F606W filter as follows:
\begin{equation}
E_{F606W}=L_{\star} \sum_{i} V_{\rm rel,i} \Delta t_{i}
\end{equation}
where L$_{\star}$ is the filter-specific stellar luminosity \citep[determined from the color-magnitude
diagram, assuming the star is at the mean bulge distance of 7.24 kpc;][]{clarkson2008},
$E_{F606W}$ is the integrated flare energy in the F606W filter, and $\Delta t_{i}$ 
is the exposure time of the $i$th time bin in the flare, and $V_{\rm rel,i}$ is
the relative flux in the F606W filter. 
We do not use photometric parallaxes from the color and apparent magnitude of the stars because
of concerns about binarity; see \S 4.2.
The zero point of the F606W filter was calculated as 2.87$\times$10$^{-9}$ erg cm$^{-2}$
s$^{-1}$ \AA$^{-1}$ using the flux calibrated spectrum of Vega \citep{calspec}.
Sequential V-band observations are 7.92 minutes
apart (start time to start time), followed typically by a gap due to
spacecraft occultation; this leads to clumping in the histogram of elapsed times
of the flare, going from 7.92 minutes to approximately 8 hours.  
The lower and upper values of the flare durations are 11 minutes and 33.9 minutes.
Since we account only for the time when the star was observed in computing flare parameters, 
these derived quantities will be lower
limits.
The error on flare energy determination is estimated by adding in quadrature the statistical
errors on each data point in the flare, and is typically $\sim$15\%, independent of stellar magnitude.  
The light curve errors are reduced for brighter stars, but the flare contrast is also smaller 
for these stars.
No error in the stellar quiescent luminosity $L_{\star}$ is included explicitly, as
the spread in distance modulus
of stars in the SWEEPS field is $\sim \pm$1 kpc from the mean bulge distance of 7.24 kpc
\citep{clarkson2008}, and any deviation of the star from the mean bulge isochrone (from which
the luminosity is estimated) imparts additional uncertainties.

\begin{figure}
\includegraphics[scale=0.5,angle=90]{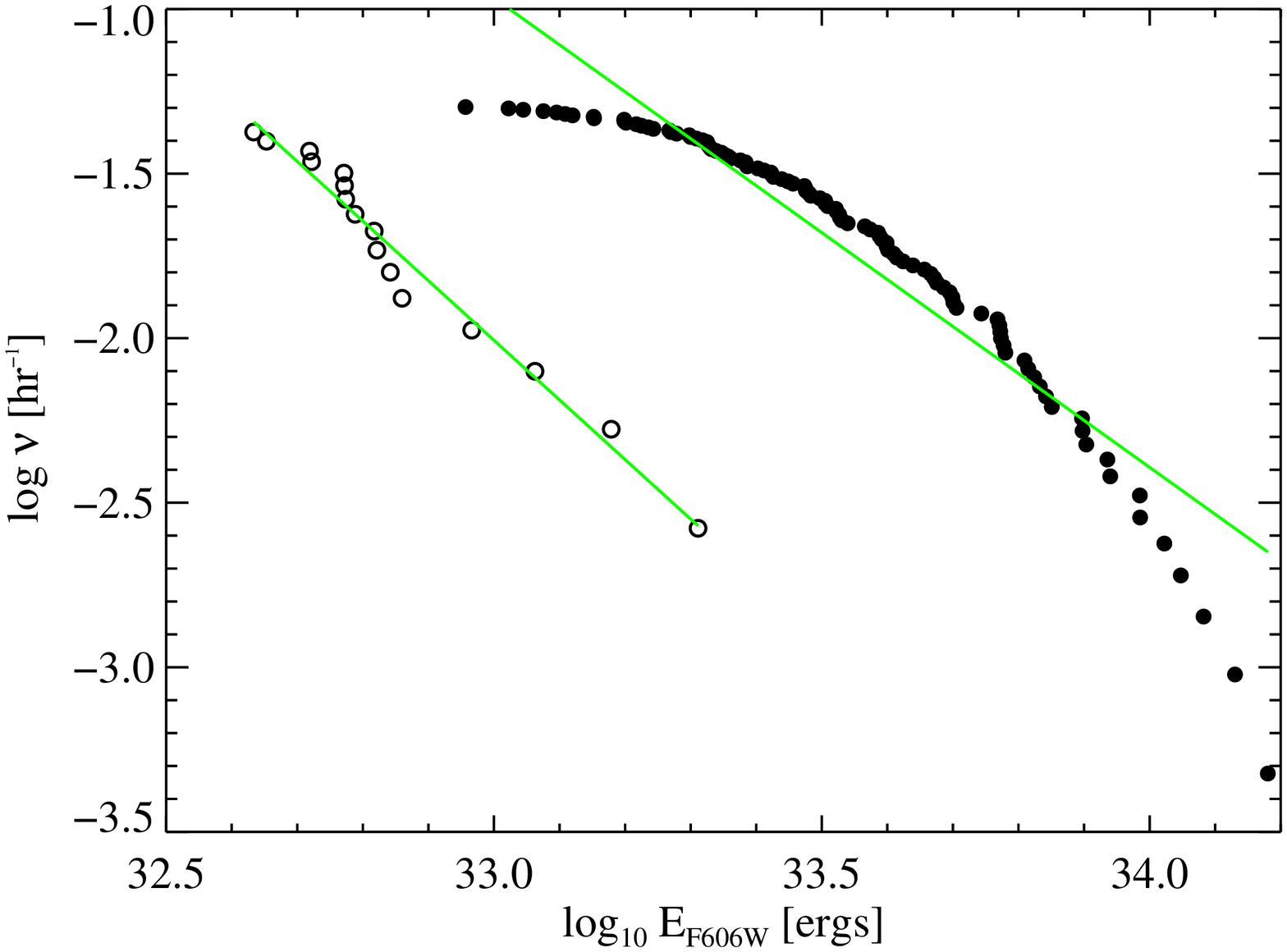}
\includegraphics[scale=0.5,angle=90]{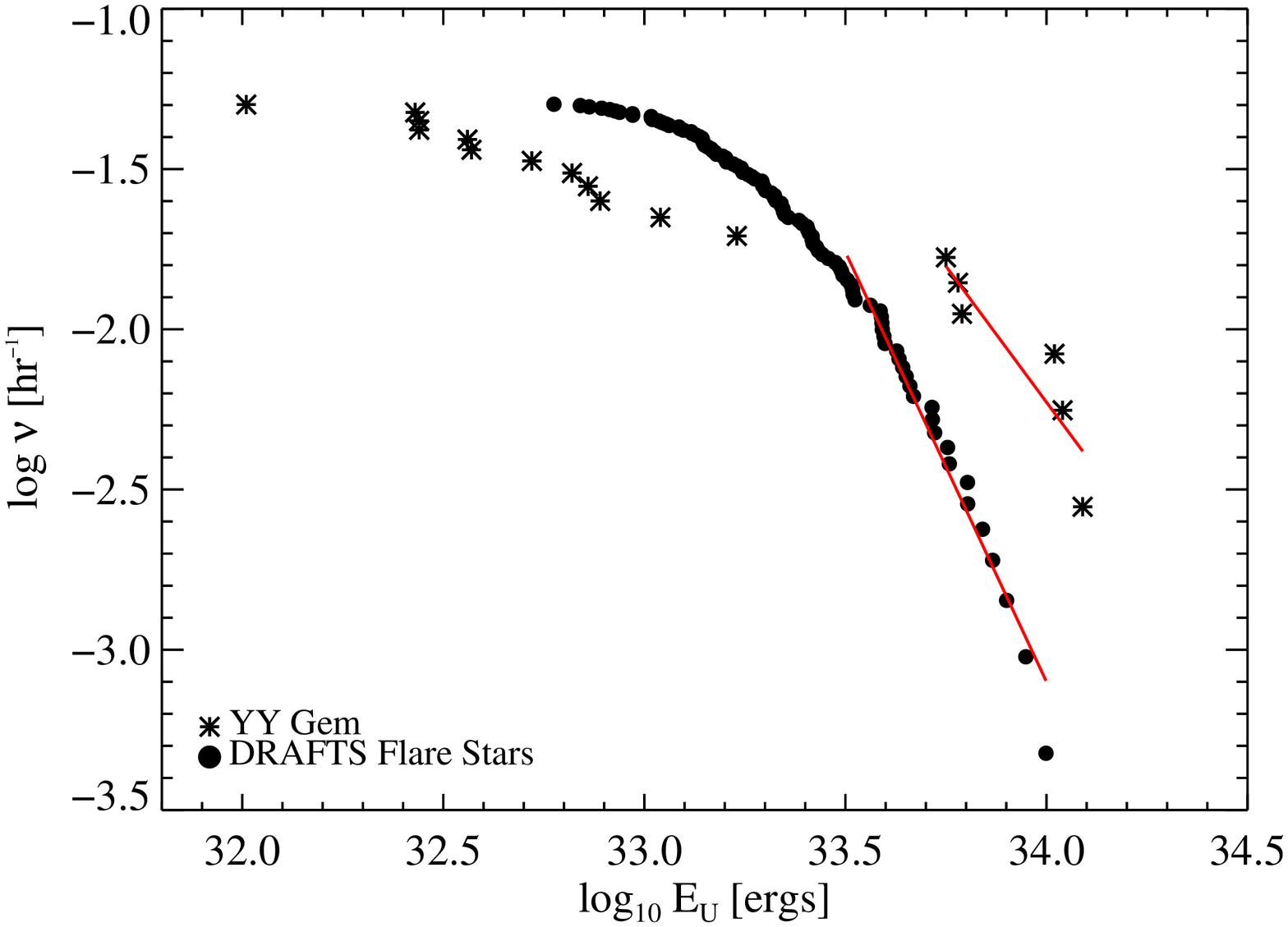}
\caption{{\it (top)} Frequency of stellar flares as a function of
flare energy, for flares occurring on non-variable stars (open circles)
and variable stars (filled circles).
The computation of quantities is described
in \S 4.1.  Green lines indicate fits to the flare frequency distributions, the parameters 
of which are listed in Table~\ref{tbl:ffds}.
{\it (bottom)} Flare frequency distribution of variable stars after conversion to an equivalent
U-band flare energy (filled circles), and compared to the flare frequency distribution of
the well-studied cool dwarf binary YY~Gem (asterisks), using the observations of \citet{lacy1976}.
The red lines show fits to the distribution above an energy of 10$^{33.5}$ erg, corresponding
to a break in the distribution of flare energies for YY Gem 
\citep{yygemflares} and to a roll over in the distribution of DRAFTS variable stars.
\label{fig:ffd}
}
\end{figure}

We considered the distributions of energies of the flares detected
on non-variable and variable stars.
\citet{kowalski2009} demonstrated that the flare frequency distribution of a large
number of stars at lower cadence is consistent with higher cadence observations
of a single star.  We calculate a cumulative flare frequency distribution by determining $\nu$, given the number of flares observed 
with a minimum energy $E$, where $\nu=N/\tau$, $N$ being the number of flares (106 and 16 for
variable and non-variable stars, respectively) and $\tau$ the total monitoring time.
The total monitoring time $\tau$ is calculated as \\
\begin{equation}
\tau = N_{s} T_{\rm mon}
\end{equation}
where $N_{s}$ is the number of flaring stars considered in each category (89 and 16 for variable
and non-variable samples, respectively), and $T_{\rm mon}$ is the
monitoring time per star (for the V band data only, this is 0.9848 days when adding up exposures in this filter
over the course of the 6.96 day observation).
The flare frequency distribution can then be plotted as $\nu$ versus $E$, and the slope of the distribution determined
from a fit to the trend in log space, using the functional form $\log \nu = \alpha + \beta \log E$ \citep{lacy1976}.
The top panel of Figure ~\ref{fig:ffd} shows the flare frequency distribution of variable and non-variable flaring stars in the F606W filter.
The derived intercepts, and slopes and errors are listed in Table~\ref{tbl:ffds}.
The slopes are the same to within the uncertainties, although it is apparent that the
distribution for DRAFTS variable stars extends to higher energies (by a factor of 8) and exhibits a turnover 
near 10$^{33.5}$ erg.
In order to facilitate comparison with other flare surveys, we convert from F606W to
U band.
We related the energy in the F606W band to that in the Johnson V band
by the ratio of the FWHM of the two filters (850 \AA\ and 2324 \AA\ for V band
and F606W, respectively), ignoring the fact that the central wavelength of the Johnson V filter
lies blueward of F606W by $\approx$400 \AA, then use the \citet{lacy1976}
relation between E$_{V}$ and E$_{U}$ (E$_{U}$=1.8$\times$E$_{V}$)
to estimate the energy of the flares in the U band (assuming that the same relation established for
flares on M stars holds for these flares as well).
The bottom plot of Figure~\ref{fig:ffd} shows a comparison between the flare frequency distribution
of the variable flaring stars in the DRAFTS sample with the nearby cool binary YY~Gem.  The
flare frequency distribution is flatter for YY~Gem than for the DRAFTS stars.
Table~\ref{tbl:ffds} lists the slopes derived from 
flares on the two sub-groups of variable stars and non-variable stars and YY~Gem.


For each flaring star, we calculate the average energy loss rate due to flares, by 
summing the radiated energy of all the detected flares for a given star, and dividing 
by the monitoring
time.  In general the errors on this derived quantity tracks the errors for the
individual flare energies, except for the case of stars with multiple flares per star,
where the error can be lower, about 10\%.
This is plotted in 
Figure~\ref{fig:lbol_eu} for the DRAFTS stars, as well as the dMe stars in 
\citet{lacy1976}.
The stellar bolometric luminosity was determined by conversion using
stellar V magnitude and the bulge isochrone of \citet{sahu2006a}. 
The bolometric luminosity for the stars in the \citet{lacy1976} sample
was taken from \citet{nlds}.
An estimate of our detection limit is shown in blue for a flare consisting of two epochs
at 3.8$\sigma$ (chosen because of the cutoff value of $\phi_{VV}=14.5$, $\sqrt{14.5}\approx3.8$).
The deviation per epoch was then used to estimate the limiting relative flux.
The higher energy flares dominate the energy budget for high luminosity stars \citep{lacy1976}.
The trend of average energy loss rate versus stellar bolometric luminosity was computed
for the ensemble of DRAFTS flaring stars, as well as the two sub-groups; these fits are
also plotted in Figure~\ref{fig:lbol_eu} and tabulated in Table~\ref{tbl:ffds}.
\citet{lacy1976} presented a strong trend of the average flare energy loss rate of dMe stars with quiescent
U band luminosity, whereas here we find a trend with quiescent bolometric luminosity.


\subsection{The Nature of the Flaring Stars}
Table~\ref{tbl:stars} in the Appendix lists the properties of the flaring stars.
Of the 214,181 stars in the non-variable sample, 16 stars flared.  Of the
1837 variable stars with successful detrending, 89 flared, giving a flaring rate among non-variable stars and
variable stars of 0.007\% and 4.8\%, respectively.  Thus, the frequency of flaring
among the variable stars exceeds that among the non-variable stars by a factor of 
$\approx$700.  
Variable stars also exhibit multiple flares per star (average of 1.25 flares per star),
in contrast to the non-flaring stars, which only had one flare per star.
We find that 71\% of the flaring stars exhibited only one flare during the 7 day
observing window.  The remainder of the flaring stars showed multiple flares, with the largest
at five flares per star.  

\begin{figure}
\includegraphics[scale=0.48,angle=90]{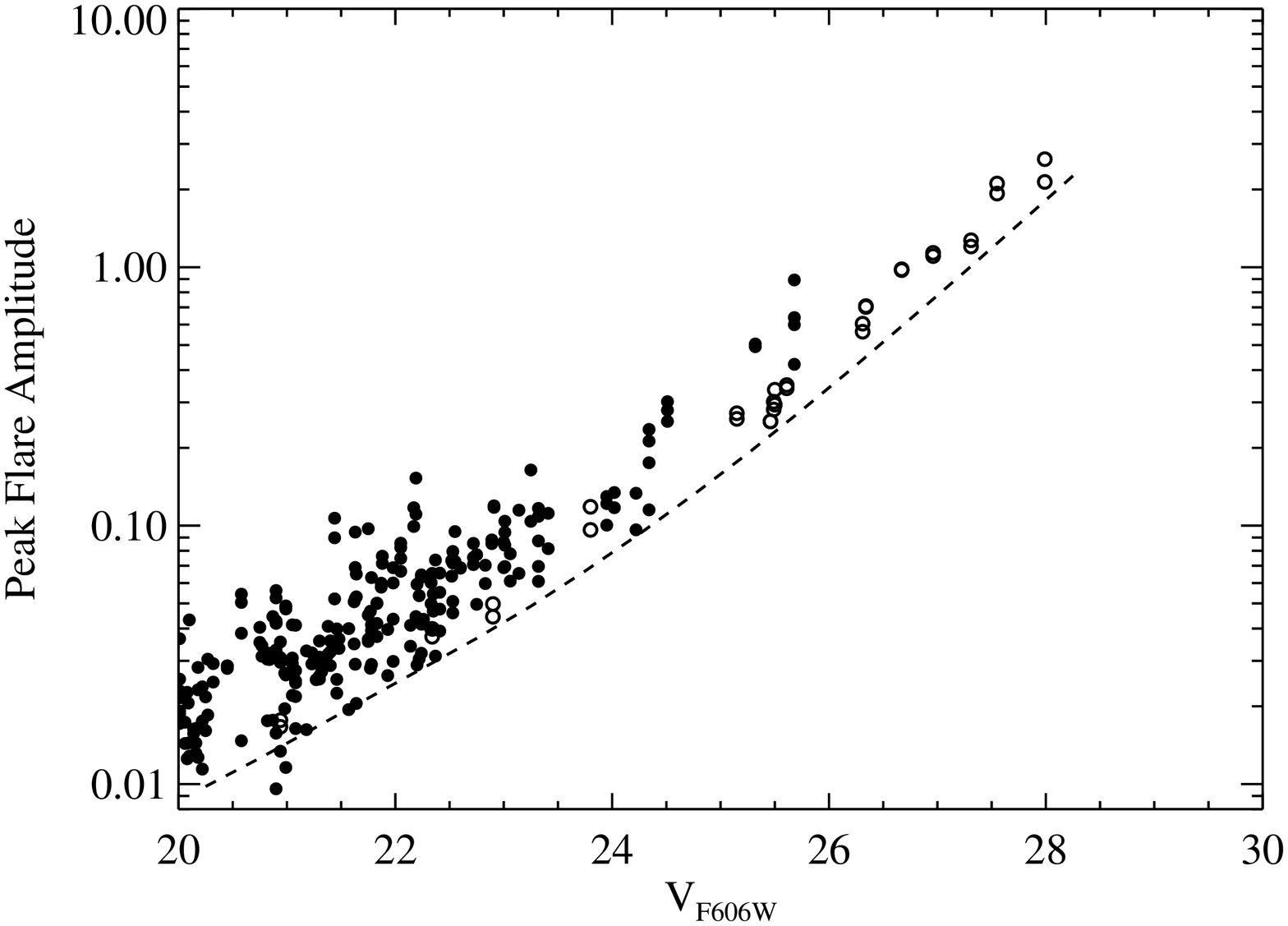}
\includegraphics[scale=0.48,angle=90]{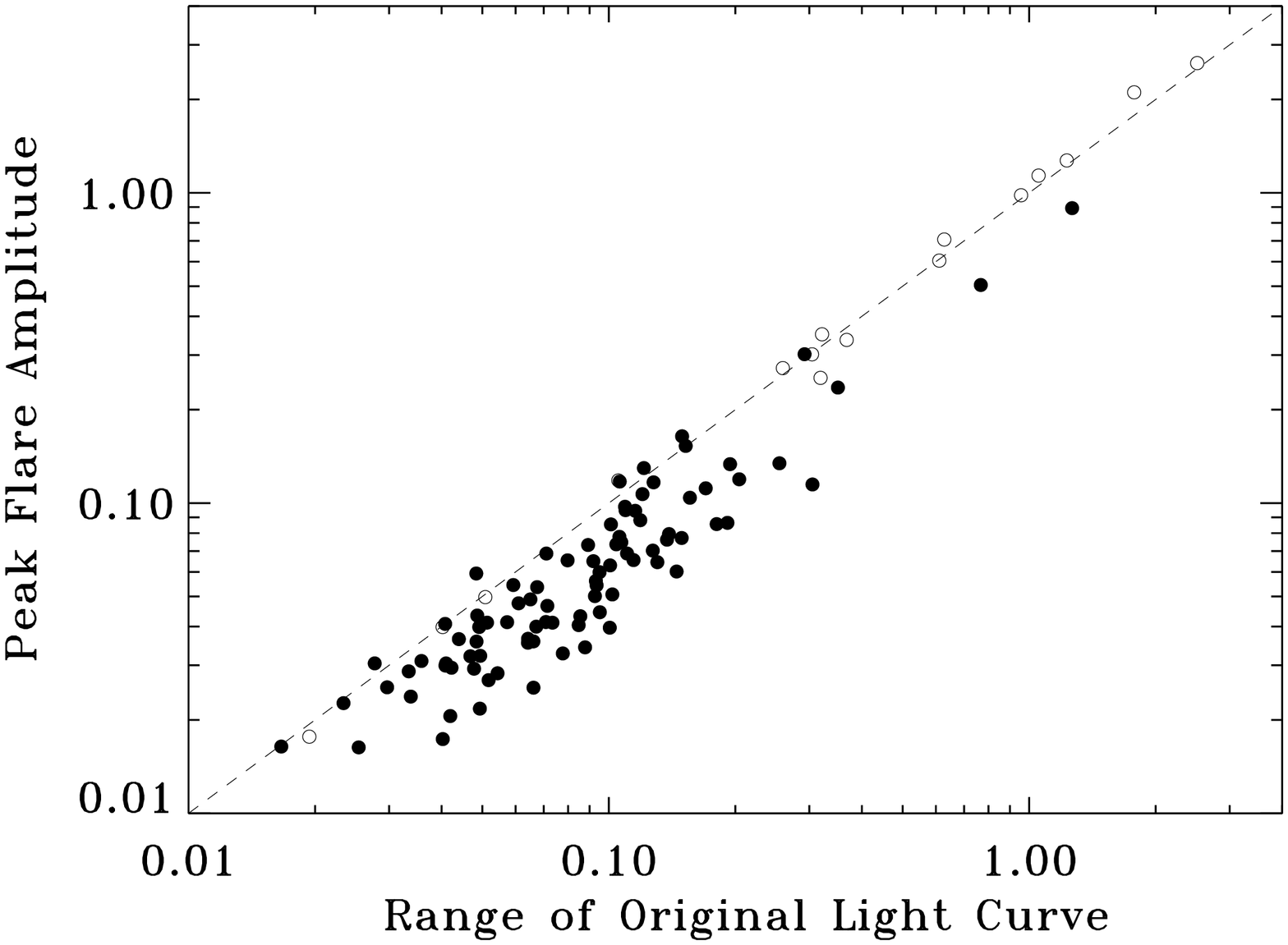}
\caption{ {\it (top)} Plot of peak flare amplitude as a function of V$_{\rm F606W}$ of the star, for
flares identified through selective filtering of the time series. Filled circles represent
flares on variable stars, and open circles flares on non-variable stars.  The dotted line represents
the noise level given by Poisson statistics for a star of a given magnitude. 
In general, the statistics on the brighter stars enabled identification of flares with 
increases of a few tens of percent, while for the fainter stars only larger variations
could be detected.
{\it (bottom)} Plot of the peak flare amplitude versus the range of the star's light curve, defined as the difference in amplitude between
the 95th and 5th percentiles, for all flaring stars.  Symbols are as in the 
top panel.  The dashed line indicates a line of equality. For stars with multiple flares, the peak flare amplitude
is the amplitude of the largest flare.  There is strong correlation between the range of underlying variability and flare amplitudes
in this sample of stars.
\label{fig:vrelvmag}}
\end{figure}

We are limited by statistics for the faint stars in our sample.  As indicated in Figure~\ref{fig:varnonvar},
past a V magnitude of about 25, the number of stars returned as variable
drops by roughly a factor of two compared to the brighter stars, and the number of non-variable stars
in this same magnitude range increases by a similar factor compared to brighter stars.  
 If we consider only stars brighter than V=25, the flare rate among variable and non-variable
stars is 87/1374 (6.3\%) and 4/94,053 (0.004\%), respectively, and the flaring frequency 
amongst these brighter variable stars exceeds that of similarly bright non-variable stars by a factor of 1600.
The non-variable
stars tend to lie at the faint end, which suggests that
the signal-to-noise of the data are limiting our ability to see underlying variations. 
These stars may be exhibiting variability at the same relative level as the brighter stars,
but our ability to diagnose this is compromised due to statistics.
Figure~\ref{fig:vrelvmag}
displays the distribution of the relative flux at the peak of the flare against
V magnitude.
Flares with an increase in flux of a few percent to a few tens of percent 
can be found for the stars at the bright end 
of the sample, and the flux increase required for a flare identification
becomes larger as the stars become intrinsically fainter.
This is a reflection of the 
statistics needed to identify flares and other longer temporal trends, which are better
constrained for the brighter stars in the sample.

There is a 
marked increase in the frequency of flaring amongst photometrically variable stars 
compared to the non-variable stars in our sample.
Of the flaring stars brighter than V=25, 96\% exhibit some kind of variability,
as gauged by the $\chi_{\nu}^{2}$ statistic. 
The bottom panel of Figure~\ref{fig:vrelvmag} 	quantifies this further, using the
``range'' of the flare light curve, taken as the difference between the values in the 95th and 5th
percentile in the original light curve (before any detrending has taken place),
and the peak flare amplitude for that star \citep{keplerref}.    For stars with multiple flares, this represents
the largest amplitude flare seen.  The non-variable stars have both the highest range and the
largest flare amplitude, both likely reflecting the lower signal-to-noise of these generally fainter
objects.

\begin{figure}
\includegraphics[scale=0.6]{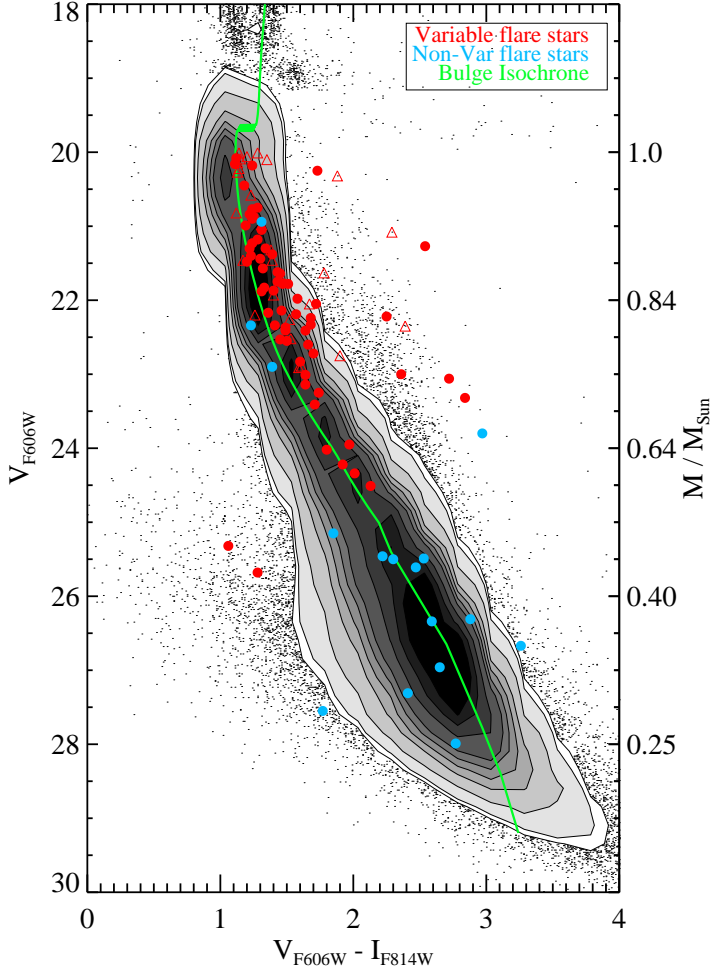}
\caption{Color-magnitude diagram (CMD) for all stars in the SWEEPS dataset.  Small black
dots give location of
individual stars where density of points is low; otherwise shaded contours are displayed.  
The green line gives a bulge isochrone appropriate for a 10 GY stellar population,
with distance modulus $(m-M)_{0}$=14.3, foreground extinction $E(B-V)=0.64$, solar metallicity,
and [$\alpha$/Fe]=0.3.  Red symbols display the 
locus of variable flare stars on the CMD, while blue circles illustrate the location
of non-variable flare stars. Regularly variable stars are indicated with a filled circle,
while the irregularly variable ones are shown with a hollow triangle.
The axis on the right side of the figure gives the appropriate mapping between luminosity and mass, using
the bulge isochrone in green.
\label{fig:bulgecmd}}
\end{figure}

Figure~\ref{fig:bulgecmd} displays the locus of the flaring stars on a color-magnitude diagram,
along with the other stars in the SWEEPS field.  In general the flaring stars lie near
the bulge isochrone, but there are several outliers, and the flaring variable stars tend to lie
systematically to the right of the bulge isochrone.  This could be due to one of three
factors: (1) young disk stars; (2) binary stars within the bulge;
or (3) stars with an enhanced metallicity.
The bulge field analyzed in this paper lies at low galactic latitude ($b$=$-2.65^{\circ}$),
and so we may 
encounter a substantial fraction of young disk stars.
Note that we cannot exclude an enhanced  metallicity as an explanation for the offset of the flaring stars
from the bulge isochrone, but we can examine the other two possibilities.

\begin{figure}
\includegraphics[bb=120 280 540 590,clip]{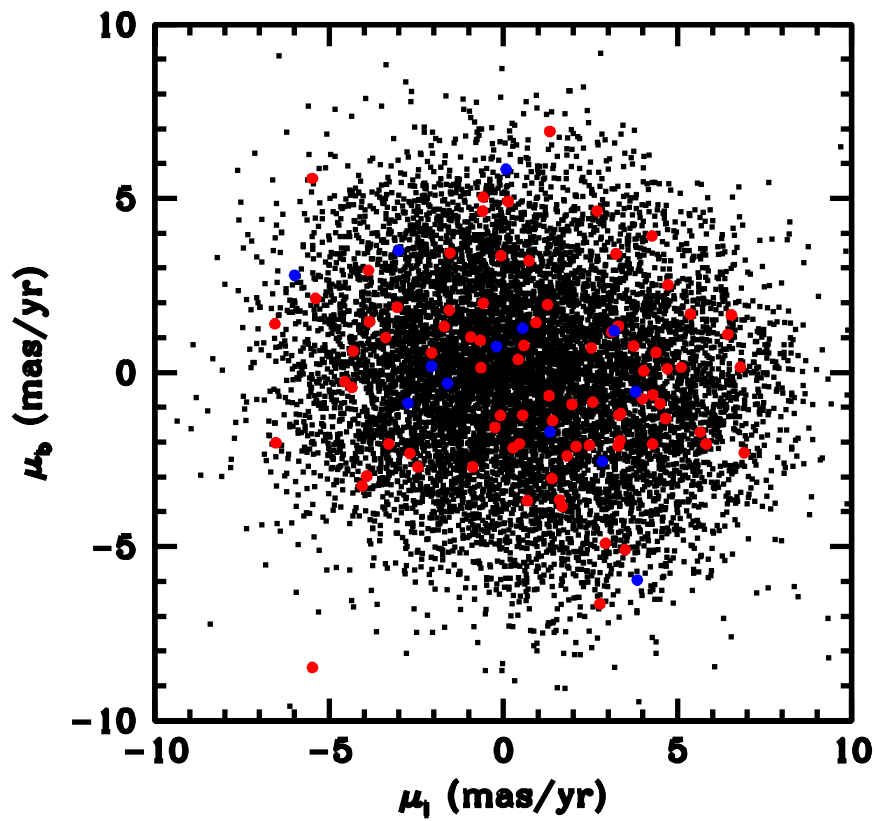}
\caption{Distribution of proper motions in galactic latitude ($\mu_{l}$) and longitude ($\mu_{b}$)
for stars in the
SWEEPS field.  The black dots show the locations of 20,000 random stars in the SWEEPS field,
while the red and blue points show the values of variable and non-variable flaring stars, respectively,
which had proper motion measurements \citep[as described in][]{clarkson2008}.
There does not appear to be any tendency for the flaring stars to have a different distribution
than the bulge stars.
\label{fig:pm}
}
\end{figure}

We make use of the results of \citet{clarkson2008}, who
computed stellar proper motions for the stars in this field using a second epoch
of data in 2006.  They were able to measure proper motions for more than 180,000
objects, and kinematically separate bulge stars from disk stars.
They determined that the fraction of disk stars in the SWEEPS field
is approximately 14\% for stars brighter than I$\sim$24.
Of the 105 flaring stars identified in the current paper, proper motions are
available for 83/89 variable stars and 14/16 non-variable stars.  Figure~\ref{fig:pm} shows the distribution of proper
motions for the flaring stars, as well as a random distribution of 20,000
stars in the SWEEPS field. The flare star distribution is very similar to
the general distribution of stars in the SWEEPS field, with no preference
for being in the disk.  The proper motion constraints are consistent with the flaring stars
being at the distance of the bulge.

\begin{figure}
\includegraphics[scale=0.7,angle=90]{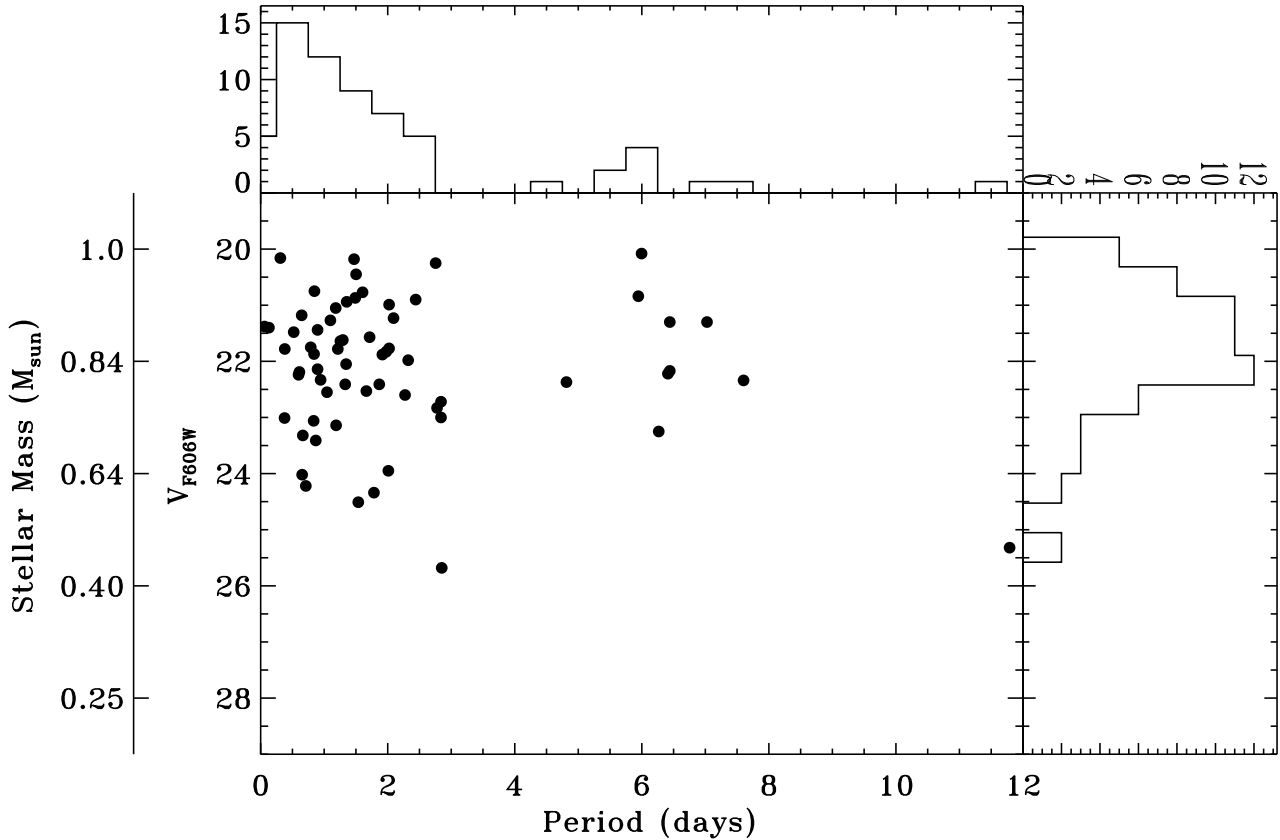}
\caption{Star's V$_{\rm F606W}$ versus period for the 63 flaring
stars exhibiting regular periodic variations. 
Most (84\%)
of these flaring stars have periods less than 3 days.
The relative number of variable stars drops for V$>$24.5 (as indicated from 
Figure~\ref{fig:varnonvar}) reflecting our inability to establish these
underlying variations.  For brighter stars there does not appear to be any bias in 
the stellar brightness versus period.
The conversion between stellar magnitude and mass is shown on the axis to the left,
using the bulge isochrone of \citet{sahu2006a}; this is only appropriate for single stars.
\label{fig:periods}}
\end{figure}

Because a large fraction of the flares detected occurred on variable stars 
(89 out of 105 stars),
we investigated further the properties of the underlying variability.
Of the 89 variable stars, 63 exhibited regular periodic variations.
We note that the periods are returned from an automatic fitting of the light curves; 
they correspond to the peak frequency of the Lomb Normalized Periodogram and are used in a 
utilitarian manner in \S 3.1 to detrend the light curves.
This analysis doesn't examine whether the period is significant, nor does it attempt to address 
biases which can creep into periodogram analyses (e.g. from red noise or period aliasing).
Also note that irregularly variable stars may have photometric periods which are statistically
significant; the detrending process in section 3.1 concentrates on which
of the two methods (given by equations 3 and 4, respectively) can better detrend the
light curves.
A plot of the period distribution of the flaring stars which exhibit
regular variations shows that 53 out of 63 have periods $\lesssim$ 3 days (Figure~\ref{fig:periods}).
This is $\approx 1/2$ the monitoring time of 6.96 days and may explain the apparent
drop-off in the period distribution past 3 days.
We note that three objects have apparent photometric periods in excess of the 6.96 day 
monitoring time, and are likely spurious. We concentrate on the objects having photometric periods
less than 3 days as these are likely the most reliable, but they are subject to the caveats described above.
Regular modulation at these timescales (for those less than about 3 days)
may be due to rapid rotation of spotted young single stars. 
However, the proper motion constraints appear to rule this out.
Another possibility is that of magnetically active (spotted) binaries (e.g. RS CVn, 
BY Dra, or W UMa binaries). 
Most of the signposts of youth 
(fast rotation, enhanced magnetic activity as evidenced by H$\alpha$ emission, X-ray
emission, starspots, flaring) are also present in active binaries, as the tidal locking
enforces fast rotation and consequent magnetic activity signatures
at old age.
Note that active binaries are typically detected via their enhanced chromospheric and/or  X-ray emission in old stellar populations
in globular clusters and old open clusters
\citep{bassa2004,vdb2004,kashyap2006},
so if tidally locked, the stars could still be at the distance
and age of the bulge population.  
As the binary fraction in the bulge is uncertain \citep{clarkson2008}, and the proper motion measurements
seem to indicate that the flaring stars are bulge members, the scenario that
these flaring stars are detached magnetically active binaries is likely.
Spectroscopic observations would be needed to confirm or refute the binary nature
of the flaring stars in favor of youth.  

We can estimate the number of active binaries expected in this field 
from the disk,
using the formalism of \citet{tinney1993}, as discussed in
\cite{koenig2008}. We calculate an effective volume V$_{\rm eff}$ which
takes into account spatial inhomogeneities probed by our sightline,
using an exponential fall-off with increasing distance from the galactic
plane.  This V$_{\rm eff}$ is  \\
\begin{equation}
V_{\rm eff} = \Omega \left( \frac{h}{\sin |b|} \right)^{3} \left[ {2-(\xi^2+2\xi+2)*exp(-\xi)} \right],
\end{equation}
where $\xi=d \sin|b|/h$, $h$ is the exponential scale height, $|b|$ the absolute value
of the galactic latitude of the pointing (here $2.65^\circ$), $\Omega$ the solid angle subtended by the 202"$\times$202" arcsec field
of view of the ACS observations, and $d$ is the mean-bulge distance, taken
to be 7.24 kpc \citep[from][]{clarkson2008}.
We use the space density of active binaries from near-field
measurements, with $n_{AB}=$3.7$\times$10$^{-5}$ stars pc$^{-3}$
\citep{favata1995} to determine the number of active binaries from the disk
we would expect to see, using $N_{AB}=n_{AB} V_{\rm eff}$, and find 
that $N_{AB}$ is between 1 and 2. This calculation shows that the expected number
of disk active binaries is low, and cannot account for the total number of flaring
objects, consistent with other constraints suggesting the bulge nature of the DRAFTS binaries.

Under the assumption that the variable flaring stars in our sample are all active binaries,
we can 
take a uniform distribution through the volume
of the bulge sampled in this exposure to constrain the space density of active binaries in the bulge. 
According to \citet{clarkson2008}
the spread in distance modulus of the SWEEPS field is about 1 kpc from the mean bulge distance of 7.24 kpc.
For a volume using the field of view translated into distance at the near and far end of this distance spread and height of 2 kpc,
we calculate a space density of active binaries of 8.8$\times$10$^{-4}$ stars pc$^{-3}$, or an enhancement of roughly
20 compared to the value found locally by \citet{favata1995}.
The disparity between the elapsed time covered in the study (6.95 days) and the monitoring time
per star in the F606W band (0.98 days) may have led us to miss flaring behavior on other
variable stars if the flares occurred during occultations or I814W observations.  The number of
variable stars is roughly a factor of 20 larger than the subset of flaring variable stars, 
so this explanation would not account for all variable stars having flares within a seven day period.
It does suggest that the enhancement of roughly 20 in the space density of active binaries in the bulge
compared to the local value
may be a lower limit.  Stellar densities are expected to be higher in the bulge compared to
the value locally, and this may lead to an enhancement in the binary fraction.
The study presented in this paper, while not specifically tuned to finding binaries, may be an indirect way to
detect the presence of magnetically active binaries through sensitive flare searches and thus place a
quantitative constraint on binary fraction in stellar associations like the Galactic bulge and old open clusters.

\section{Discussion}
The sample of stars surveyed for flares in this paper is markedly different
from the types of stars and wavelength ranges usually surveyed for stellar flares.
Because of the large amount of contrast with underlying coronal emission and
the impulsive nature of flares, many studies of flares 
have concentrated on the high energy (extreme ultraviolet to X-ray)
region \citep{wolk2005,stelzer2000,audard2000,ostenbrown1999}
to characterize flares on young stars, active stars and binaries.
Due to the intrinsically red light of M stars in quiescence, 
their blue flares are easily studied
in the optical \citep{lacy1976,kowalski2009,hilton2010}.  More recently,
the long stellar time series returned by the Kepler mission have been explored
in the context of white light flares on stars later than K0 spectral
type \citep{keplerref} for long duration flares lasting more than 1.5 hours.
The high energy measurements have been flux- and therefore distance-limited,
and the optical flare studies have preferentially focused on younger stars, due
to the known decline of magnetic activity and flaring with age (and the correlation between
stellar age and increasing distance from the Galactic plane).
This study is the first to cover a wide range of cool stellar spectral types
at optical wavelengths, while sampling an older population of stars.

Our supposition is that our sample of flaring bulge stars is dominated by a 
population of cool close binaries displaying magnetic activity through transient
flaring activity.  
The magnetic activity of such close binaries is usually studied through their coronal 
and chromospheric time-averaged properties; multi-wavelength flare campaigns 
to study the properties of flares on solar neighborhood binaries 
\citep{stern1992,osten2002,osten2004} have concentrated on 
ultraviolet, X-ray, and radio measurements.
As discussed by \citet{henry1996}, there are very few observations of optical flaring on nearby 
active detached binaries with
G and K primaries -- by far the most attention has been on
optical studies of late K/M dwarf binaries such as YY Gem \citep{yygemflares}.
One notable exception is a large optical flare on the binary HR~1099 (K1IV+G5V) observed  by
both \citet{zhang1990} and \citet{hh1991}, with peak enhancements noted in the V band
(at different times)
of 0.18 and 0.42 magnitudes, respectively.
\citet{m1992} and \citet{henry1996} determined the optical flare rate for 
the nearby active binary II~Peg (K2 IV+dM) to be near 0.2 flares hr$^{-1}$ using different data
sets, although \citet{henry1996} noted variation in the flare rate.
\citet{rucinski1985} and references therein mention that a few flares
have been seen in contact binaries of the W UMa type (containing magnetically active 
stars), and they appear to follow the relations of M dwarf flare stars.
Such observations do not generally suffice to compute a flare frequency distribution
with energy or other flare statistics to which we can compare our results, and so
we instead look to the statistics on optical M dwarf flares for comparison.

\begin{figure}
\includegraphics[scale=0.7,angle=90]{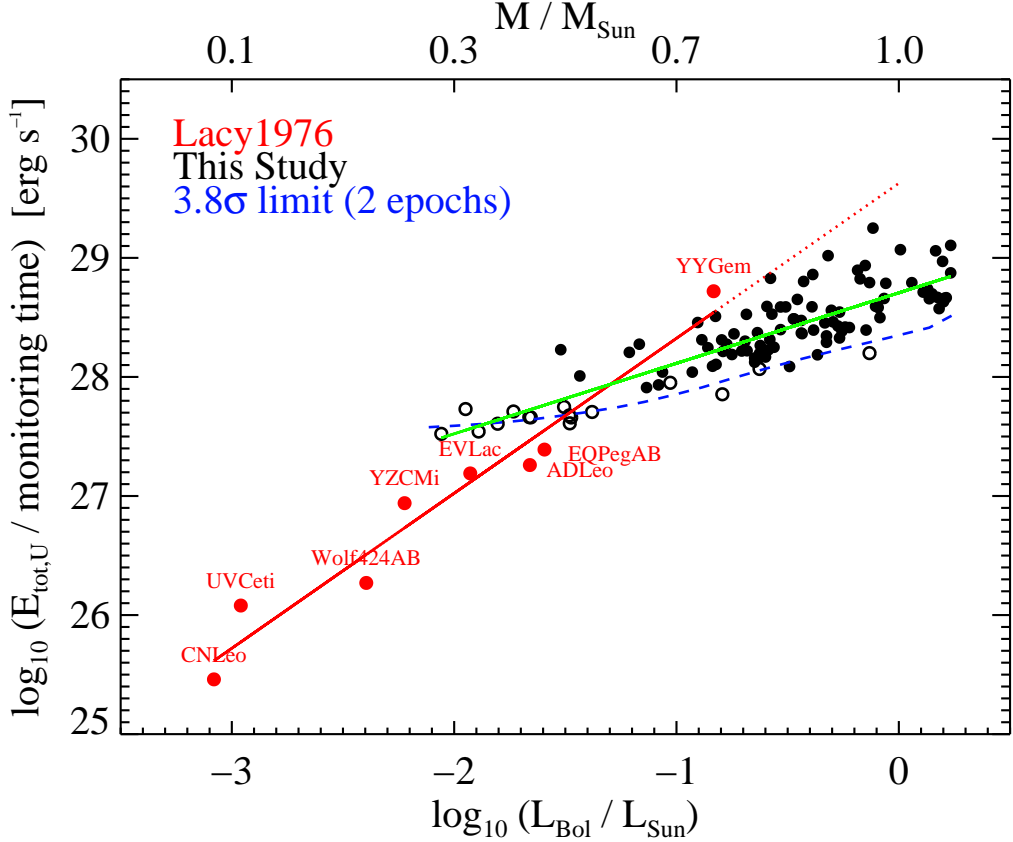}
\caption{The energy loss rate due to flares plotted against stellar bolometric luminosity, for
each star identified as a flare star.  
The calculation of quantities is discussed
in \S 4.1.
Filled black circles represent results for variable DRAFTS flare stars, while open circles
show results for non-variable DRAFTS flare stars.  
Red filled circles are results from \citet{lacy1976} for a sample of M dwarf flare stars.
The blue dashed line indicates the approximate limit based on 3.8 $\sigma$ detections at
two epochs, which gives the cutoff value $\phi_{VV}$ of 14.5.  The deviation per epoch was used
to get a limiting relative flux. 
A fit to the trend of flare energy loss rate vs stellar luminosity is shown 
for all DRAFTS stars in green, and the red line illustrates a fit to the dMe
stars in \citet{lacy1976}, extrapolated to $L_{\rm bol}$=L$_{\rm sun}$.
The axis at the top of the plot lists conversion between stellar bolometric light
and mass using the bulge isochrone, and is only appropriate for single stars.
The average energy loss rate for the flaring stars identified here far exceeds all of the
M dwarf flare stars studied in \citet{lacy1976} except for the close binary YY~Gem, although the steeper trend of the dMe stars would imply
a larger energy loss rate at $L_{\rm bol}$=L$_{\rm sun}$.
\label{fig:lbol_eu}}
\end{figure}

As noted  in \S 4.1, the flare frequency distributions of the non-variable and variable DRAFTS stars
contain several notable differences.  The DRAFTS variable flaring stars have both more numerous
and more energetic flares than the non-variable flaring stars, with maximum energy roughly a factor of 10 larger.
The slopes of the distributions are similar to within the uncertainties, however the most
energetic flares on the DRAFTS variable flaring stars have a much steeper slope of $-2.68$ above 10$^{33.5}$ erg
(see Table~\ref{tbl:ffds}.)
The bottom panel of Figure~\ref{fig:ffd} compares the flare frequency distribution with energy of the ensemble 
of stellar flares from DRAFTS variable stars against that determined by \citet{lacy1976}
for optical flares on the M dwarf binary YY~Gem. This binary, composed of
two M1 stars \citep{starspots} was chosen because of
its general level of enhanced magnetic activity, as evidenced by starspots, chromospheric and coronal emission,
and a high flaring rate. It also has the advantage of being relatively well-observed 
in its flare properties.
The largest stellar flare energies from YY Gem and the DRAFTS variable flare stars are similar, assuming that the same conversion
applies; the study of YY Gem is clearly sensitive to lower energy events.
The distribution of flares with energy for YY Gem has an apparent break around 10$^{33.5}$
erg, which is also near where the distribution of flare energies in the DRAFTS variable
stars shows evidence of a rollover.  Table~\ref{tbl:ffds} shows the slopes
of the distributions computed for this high energy end.  The DRAFTS flaring variable stars
show a much steeper trend.
The few optical flares
seen on nearby active binaries have exhibited flare energies that can approach
10$^{6}$ times the largest solar flare energies of 10$^{32}$ ergs \citep{hh1991},
although the evolved nature of the flaring binary member(s) may play a role
in these extremes.  
\citet{yygemflares} noted that the binary nature of YY~Gem may be responsible for the
change in the flare frequency distribution above 10$^{33.5}$ erg, as the increased volumes allow for
more energy to be stored and released during flares.
Figure~\ref{fig:lbol_eu}  plots the average energy loss rate due to flares
for the flaring stars in our sample and the sample of M dwarf flare stars
studied by \citet{lacy1976}. 
YY Gem has the 
highest average energy loss rate of the dMe stars considered in \citet{lacy1976}.
It even exceeds the DRAFTS stars at the same total light.
\citet{ferreira1998} pointed out that such energetic flares in cool binaries
require strong magnetic fields on at least one of the stars, and possibly also a large volume.

\citet{lacy1976} found that the luminous M dwarfs emit more energy in flares
than the faint M dwarfs, even when one accounts for the smaller ``flare visibility''
on the more luminous stars.  
We find that the energy loss rate of the non-variable stars tends to be smaller than
variable stars, even taking into account the general faintness of the former compared
with latter, which biases our flare detection to larger flare events (fits to the subsets
and the entire set of DRAFTS flaring stars are given in Table~\ref{tbl:ffds}).
This confirms the results shown in Figure~\ref{fig:ffd} which demonstrate more 
frequent and energetic flaring for the DRAFTS variable flare stars.
The dMe stars in \citet{lacy1976} show a steeper trend of flare energy 
loss rate vs. luminosity than the DRAFTS flaring stars in Figure~\ref{fig:lbol_eu}, even though 
the average energy loss rate of the DRAFTS flaring stars can be up to 30 times higher.
The energy loss rate of these trends at the solar value suggests roughly an order of magnitude higher
flare energy loss rate for the dMe stars than for the DRAFTS flaring stars.
It is thought that the relation for M dwarfs
turns over at spectral type dK since even the largest solar flares are much weaker than
what an extrapolation predicts.  
Possibilities for the different slopes may arise from the different internal structures of the dMe
stars considered in \citet{lacy1976} compared to the higher luminosity/mass DRAFTS flaring stars,
or the hypothesized binary nature of the DRAFTS variable stars.  We note that the one
dMe star in that sample which is a close binary is YY~Gem, whose average flare energy loss rate
is in the range of the DRAFTS flaring stars.  If the bulk of the DRAFTS flaring stars are
indeed binaries, this would shift the L$_{\rm bol}$/L$_{\rm sun}$ values for DRAFTS flaring stars in Figure~\ref{fig:lbol_eu}
to the left, and bring the values into better agreement with YY~Gem.

\begin{figure}
\includegraphics[scale=0.7,angle=90]{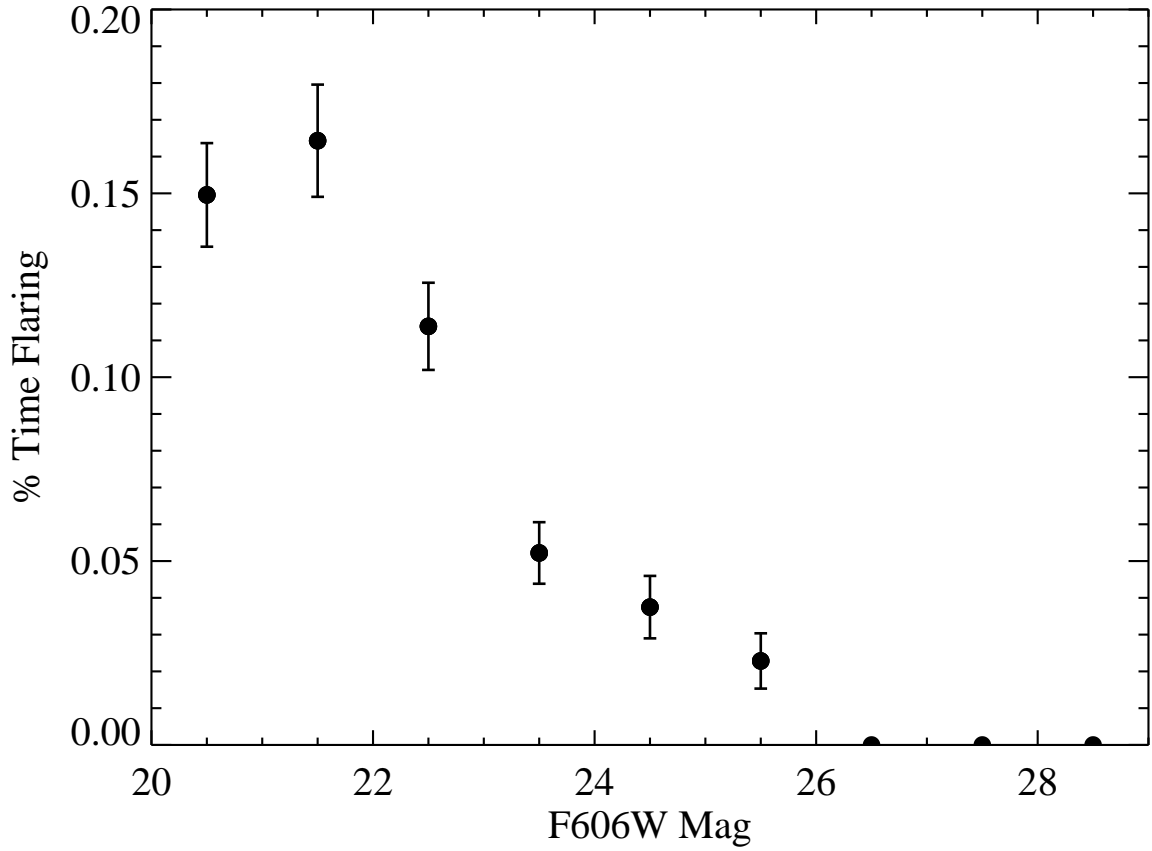}
\caption{Variation of the fraction of epochs affected by flaring against stellar
magnitude, for all variable stars.  Percentages were computed by summing the number of epochs in a given
magnitude bin which were identified as a flare by our algorithm, and dividing by the
total number of epochs in that magnitude bin. For fainter stars our ability to detect flares
becomes more difficult. 
\label{fig:fl_fraction}
}
\end{figure}

Overall, only a small fraction of stars in the sample flare, but their flare rates
are large by comparison with measures of flaring in other kinds of active stars.
The 105 flaring stars represent  0.05\% of all the bulge stars searched for flares, taking
the 14\% disk fraction of \citet{clarkson2008} and the number of stars with V$>20$. 
Figure~\ref{fig:fl_fraction} shows the variation of flaring fraction versus stellar magnitude for
the stars exhibiting underyling variability.  For each magnitude bin, the number of flaring
epochs was divided by the number of total epochs in that magnitude bin.
\citet{kowalski2009} found a maximum flare rate of M dwarfs near the galactic plane 
(at $b\sim-38^{\circ}$)
for flares with $\Delta u>$0.7 mag on stars with $u<$22
of 8 flares  hr$^{-1}$ deg$^{-2}$.
The flaring stars in the Kepler field studied by \citet{keplerref}
were of spectral type K0V and cooler, and
had much smaller peak amplitudes, at $\Delta F/F>$0.1\% generally,
but represent an approximate flaring rate of 0.03 flares hr$^{-1}$ deg$^{-2}$
(this number may actually be higher, even for the stellar range considered in the paper,
since only a subset of stars in the 105 degree$^{2}$ Kepler field are selected for monitoring).
We compute the spatial flare rate for the DRAFTS flaring stars using the effective monitoring time of the dataset (0.9848 days)
and solid angle subtended by the field of view, 0.003148 deg$^{2}$, which for 122
flares results in a flare rate of 1700 flares hr$^{-1}$ deg$^{-2}$, two orders of magnitude greater
than  that found by \citet{kowalski2009}.
The bulge is of course a much more crowded field than any Galactic disk field, with an intrinsically
larger number of stars per square degree, which will contribute to this enhancement.
With sensitive
photometry, this study is able to pick out much smaller flares on a larger variety of cool stars, 
with peak flare amplitudes as small as a 
few percent, which may partly explain the much higher flare rate.
The flare rate from this study can also be expressed as 1.2 flares per flaring star per day, 
which is about ten times higher than the X-ray flare rate of 0.13 flares per flaring star per day
\citet{wolk2005} determined on young single Suns in the Orion Nebula  Cluster.
\citet{ostenbrown1999} determined the extreme ultraviolet flare rate among a sample of
nearby active binary systems to be in the range 0.1-1.5 flares per day, or 0.3 flares per day
taking the entire flare sample as one.  The DRAFTS flare rate is consistent with the upper end
of this distribution, suggesting that the high precision of this optical study
may balance the higher contrast in the extreme ultraviolet wavelength observations.
These statistics demonstrate that the flaring stars found in this study 
exhibit flaring activity at the extremes found in magnetically active stars.
This is also supported by the frequency distributions, comparable to one of the most
energetic nearby flare stars, as well as by the increased average flare energy 
loss rate compared to most well-studied nearby M dwarf flare stars.

The high flaring rate has implications for current and future time-domain studies:
given the derived spatial flare rate of 1700 flares hr$^{-1}$ deg$^{-2}$,
a single 15 second frame of an exposure with the Large Synoptic Survey Telescope (LSST)
\footnote{More information can be found at \texttt{http://www.lsst.org}.}
covering 9.6 deg$^{2}$ might return as many as 68 flares with amplitudes greater than 1\%
of the quiescent luminosity.  This is the same order as the maximum of 50 flares per
LSST exposure from M dwarfs at low galactic latitudes with flares exceeding $\Delta u>$0.1 mag
\citep{hilton2010cs},
and suggests an underestimate of the contribution to the variable sky from flaring cool stars.

The existence of flaring in these old stars opens up other questions regarding the longevity
of binary systems at this age.
The traditional assumption about a decline of magnetic activity and flaring with stellar age
arises mainly from consideration of single stars \citep{skumanich1972,soderblom1982}.
For stars in a binary system close enough to have tidal interactions, 
tidal effects can cause synchronous rotation,
circularizing the orbit and coupling the stars' rotation
to the orbit.  This enforces fast rotation at old age, as discussed earlier.
The synchronization time for stars with convective envelopes is relatively short,
approximately 1 MY for near-unity mass ratios and orbital periods less than 3 days \citep{zahn1977},
and a binary with an orbital period $<$ 3 days would circularize in about 1 GY.
One can ask, though, how long such binaries can be maintained against angular momentum loss,
particularly in the present case where we may be seeing flares from 10 GY binaries.
The spin-orbit coupling which leads to enhanced magnetic activity can also result in
orbital evolution of the system.  Whereas in a single star, magnetic torques in a stellar wind
lead to angular momentum loss and subsequent spin-down with time, in a close binary 
the coupling of the spin and orbital angular momenta paradoxically leads the 
angular momentum loss from the stellar wind to result in orbital shrinkage and associated faster rotation \citep{guinan1993}.
The exact timescale is given by the binary mass ratio, the initial period, and the nature of the stellar wind
and its effect on magnetic braking.  
The angular momentum evolution of cool close binaries has been studied quantitatively 
\citep{guinan1993,stepien2011}, using stellar wind/velocity variations with time found for single stars.
Differences in these two treatments, which stem from different parameterizations of the angular momentum loss,
can lead to differences in the timescales for evolution of a cool close
binary to the contact stage or Roche Lobe overflow of up to an order of magnitude.

The evolution of close binaries in the bulge could also be affected by 
the higher number densities of stars in the bulge: a tertiary
can affect the angular momentum evolution of close binaries through Kozai cycles accompanied by
tidal friction  \citep{ft2007}.  
Another possibility for these stars, if spectroscopic confirmation shows that some of these are single,
is that they are magnetically young (by virtue of their flaring and fast rotation) but kinematically
old.  They would then be similar to some solar neighborhood chromospherically young, kinematically old
stars studied by \citet{poveda1996} and \citet{rochapinto2002}, and a plausible explanation for their existence at such old ages 
would be binary coalescence of a low mass, short period binary system into an apparently single,
rapidly rotating (and flaring) star.
Thus, while we do not know the exact nature
of the stars in our sample, their activity (as deduced by the flares detected on them) and
their 10 GY age (from proper motions consistent with the bulge) coupled with the few day
periods observed from the data suggest that they may be able to 
place constraints on the angular momentum evolution of binary stars, if they are close binaries,
or signal the formation timescales and coalescence rates of low mass stellar mergers.

\section{Summary and Conclusions}

We repurposed the Sagittarius Window Eclipsing Extrasolar Planet Search into a
Deep Rapid Archival Flare Transient Search.  Data from the nearly 7 day stare, coupled
with rigorous tests to select candidate flare events, allowed us to identify 
122 flares on 105 stars, out of a total of 216,136 stars surveyed. 
The characteristics of the flares are similar to those of 
optical flares on 
well-studied, nearby flaring stars, and suggest a high flaring rate.
Stars with underlying variability produced flares at a higher rate 
than stars with no discernible variability, by a factor of nearly 700; this factor is 1600
when considering stars between V=20 and V=25.
We interpret these underlying variations as arising from rotational modulation
due to starspots.
The flaring stars appear to be characterized by an older stellar population
which is likely binary in nature, and demonstrates the capability of
tidally locked systems to retain magnetic activity in old age.
The study expands the type of stars studied for flares in the optical band,
and suggests that future optical time-domain studies will have to contend with
a larger sample of potential flaring stars than the M dwarf flare stars usually considered.
A more in-depth study of the nature of the variable flaring stars in the present
sample will provide constraints on the nature of angular momentum evolution
in old cool, magnetically active  binary stars.

\acknowledgements

Based on data taken with the NASA/ESA Hubble Space Telescope 
obtained at the Space Telescope Science Institute (STScI). 
Support for this research was provided by NASA through grants 
AR-17767.0 and AR-17767.1 
from STScI, which is operated by the Association of Universities for Research in Astronomy (AURA), inc., under 
NASA contract NAS5-26555.
AFK and SLH acknowledge support from NSF grant 08-07205.
We thank A. C. Becker, E. J. Hilton, and B. P. Brown for useful discussions.
We also thank J. R. A. Davenport for the use of his contour routine.

\clearpage
\begin{deluxetable}{lcc}
\rotate
\tablewidth{0pt}
\tablenum{1}
\tablecolumns{3}
\tablecaption{DRAFTS Flare Search Breakdown \label{tbl:nbrs}}
\tablehead{ \colhead{Selection Criterion} & \colhead{\# in Variable Sample} &  \colhead{\# in Non-Variable Sample} }
\startdata
Stars in the field & \multicolumn{2}{c}{229,701} \\
Stars with V$<$29.5 & \multicolumn{2}{c}{229,293} \\
Stars in which $>$70\% of light curve bins have errors $<$ 3 $\sigma$& \multicolumn{2}{c}{222,657} \\
Stars with V$>$20 & \multicolumn{2}{c}{216,136} \\
$\chi^{2}_{\nu} <$1.5 (non-variable) or $\chi^{2}_{\nu} >$1.5 (variable) &1955  & 214,181 \\
Variable stars with successful detrending & 1837 & --\\
Epoch pairs with $\phi_{VV}>$14.5, V$_{\rm rel,i,j}>$0 ($j=i+1$) & 920 & 17 \\
Epoch pairs with V$_{\rm rel,i}$/$\sigma_{i} >$ 2.5 & 229 & 17 \\
Individual flare detections & 128 & 16 \\
\# flare events & 106 & 16 \\
Total \# of flaring events & \multicolumn{2}{c}{122} \\
Flaring stars & 89 & 16 \\
Total \# of flaring stars & \multicolumn{2}{c}{105} \\
\enddata
\end{deluxetable}

\begin{deluxetable}{llll}
\tablewidth{0pt}
\tablenum{2}
\tablecolumns{4}
\tablecaption{Flare Frequency Distributions \label{tbl:ffds}}
\tablehead{ \colhead{Category} & \colhead{Band} & \colhead{Intercept} &\colhead{Slope} }
\startdata
\hline
\multicolumn{4}{c}{$\log \nu = \alpha + \beta \log E$} \\
\hline
DRAFTS variable stars & F606W & 46.1 & -1.43$\pm$0.14 (N=106) \\
DRAFTS non-variable stars & F606W & 57.8 & -1.81$\pm$0.45 (N=16) \\
DRAFTS variable stars & U band & 45.8 & -1.43$\pm$0.14 (N=106) \\
YY Gem & U band & 14.5 & -0.49$\pm$0.12 (N=18) \\
DRAFTS variable stars, E$_{U,fl}>$10$^{33.5}$ erg & 88.1& U band &-2.68$\pm$0.49 (N=30) \\
YY Gem, E$_{U,fl}>$10$^{33.5}$ erg & U band & 55.5& -1.69$\pm$0.69 (N=6) \\
\hline
\multicolumn{4}{c}{$\log E_{\rm tot,U}/T_{\rm mon} = \alpha + \beta \log L_{\rm bol}/L_{\odot}$} \\
\hline
DRAFTS all stars & U band & 28.7 & 0.59$\pm$0.04  (N=122)\\
DRAFTS variable stars & U band & 28.7 & 0.53$\pm$0.05 (N=106)\\
DRAFTS non-variable stars & U band & 28.2 & 0.33$\pm$0.04  (N=16)\\
\citet{lacy1976} sample & U band & 29.6 & 1.30$\pm$0.12 (N=8)\\
\enddata
\end{deluxetable}

\section{Appendix}
Table~\ref{tbl:stars} lists the properties of the DRAFTS flaring stars: their positions,
magnitudes in the F606W and I814W filters, the type of variability, the period
of any underlying regular variability, and the range of the amplitude.

\begin{deluxetable}{llllcll}
\tablewidth{0pt}
\tablenum{A1}
\tablecolumns{7}
\tablecaption{Properties of DRAFTS Flaring Stars \label{tbl:stars}}
\tablehead{ \colhead{RA } & \colhead{Dec } & \colhead{V } & \colhead{I )} &\colhead{Variability }
& \colhead{Period} & \colhead{Range\tablenotemark{a}} \\
\colhead{ (J2000)} & \colhead{(J2000)} & \colhead{(F606W)} & \colhead{(F814W)} & \colhead{Type} & \colhead{(d)}  &\colhead{} 
}
\startdata
  17  58  53.38  &  -29  11  52.58  &    25.49  &    22.96    &  flat   &    \nodata&       0.30\\
  17  58  54.29  &  -29  10  26.85  &    20.94  &    19.63    &  flat   &    \nodata&       0.02\\
  17  58  54.38  &  -29  13  25.05  &    22.90  &    21.51    &  flat   &    \nodata&       0.05\\
  17  59   0.37  &  -29  12   6.01  &    25.15  &    23.30    &  flat   &    \nodata&       0.26\\
  17  58  59.43  &  -29  10  24.00  &    27.99  &    25.22    &  flat   &    \nodata&       2.51\\
  17  58  53.98  &  -29  10  27.79  &    25.50  &    23.20    &  flat   &    \nodata&       0.37\\
  17  58  53.54  &  -29  12   7.09  &    27.55  &    25.78    &  flat   &    \nodata&       1.78\\
  17  59   0.54  &  -29  11   0.50  &    26.96  &    24.31    &  flat   &    \nodata&       1.05\\
  17  58  58.37  &  -29  11  15.66  &    26.34  &    23.75    &  flat   &    \nodata&       0.63\\
  17  59   4.34  &  -29  12   2.30  &    25.61  &    23.14    &  flat   &    \nodata&       0.32\\
  17  59   3.17  &  -29  12  18.53  &    26.67  &    23.41    &  flat   &    \nodata&       0.96\\
  17  59   5.10  &  -29  13   7.97  &    23.80  &    20.83    &  flat   &    \nodata&       0.11\\
  17  59   2.92  &  -29  12  43.19  &    22.34  &    21.11    &  flat   &    \nodata&       0.04\\
  17  59   4.61  &  -29  13  27.75  &    25.46  &    23.24    &  flat   &    \nodata&       0.32\\
  17  59   1.35  &  -29  12  46.19  &    27.31  &    24.90    &  flat   &    \nodata&       1.23\\
  17  59   6.88  &  -29  12  37.67  &    26.31  &    23.43    &  flat   &    \nodata&       0.61\\
  17  59   0.31  &  -29  13  22.41  &    21.30  &    20.08    &  var/reg&       6.44    &   0.04\\
  17  58  54.60  &  -29  11  36.28  &    21.40  &    20.17    &  var/reg&       0.13    &   0.07\\
  17  58  57.40  &  -29  13  18.26  &    20.87  &    19.62    &  var/reg&       1.49    &   0.10\\
  17  58  59.00  &  -29  11   4.57  &    20.94  &    19.63    &  var/reg&       1.35    &   0.06\\
  17  58  58.07  &  -29  12  13.00  &    21.88  &    20.57    &  var/reg&       1.92    &   0.14\\
  17  58  54.06  &  -29  12  15.28  &    21.75  &    20.32    &  var/reg&       0.79    &   0.11\\
  17  58  54.88  &  -29  10  29.63  &    20.22  &    19.08    &var/irreg&       \nodata    &   0.03\\
  17  58  54.65  &  -29  12  11.71  &    20.18  &    18.94    &  var/reg&       1.47    &   0.05\\
  17  58  55.31  &  -29  11  28.88  &    20.82  &    19.70    &var/irreg&       \nodata    &   0.04\\
  17  58  59.25  &  -29  10  57.39  &    20.32  &    18.44    &var/irreg&       \nodata    &   0.05\\
  17  58  57.10  &  -29  12  58.73  &    20.99  &    19.80    &  var/reg&       2.02    &   0.07\\
  17  58  56.37  &  -29  12  54.06  &    22.41  &    20.77    &  var/reg&       1.33    &   0.06\\
  17  58  59.93  &  -29  11  55.12  &    20.09  &    18.94    &var/irreg&       \nodata    &   0.04\\
  17  58  53.74  &  -29  10  47.26  &    20.25  &    18.52    &  var/reg&       2.75    &   0.05\\
  17  59   0.41  &  -29  11  44.32  &    21.93  &    20.53    &var/irreg&       \nodata    &   0.10\\
  17  58  55.17  &  -29  12  15.48  &    21.57  &    20.25    &  var/reg&       1.72    &   0.07\\
  17  58  55.14  &  -29  12  48.93  &    21.30  &    19.96    &  var/reg&       7.03    &   0.05\\
  17  58  53.09  &  -29  12  50.24  &    22.75  &    20.85    &var/irreg&       \nodata    &   0.15\\
  17  58  54.20  &  -29  12  29.68  &    20.06  &    18.86    &var/irreg&       \nodata    &   0.04\\
  17  58  55.98  &  -29  12  31.97  &    21.98  &    20.40    &  var/reg&       2.32    &   0.11\\
  17  58  56.88  &  -29  11  11.69  &    22.20  &    20.94    &var/irreg&       \nodata    &   0.05\\
  17  58  57.37  &  -29  10  40.02  &    22.17  &    20.81    &  var/reg&       6.44    &   0.11\\
  17  58  57.76  &  -29  12   8.28  &    20.45  &    19.27    &  var/reg&       1.50    &   0.03\\
  17  58  56.26  &  -29  13  36.83  &    21.38  &    19.99    &  var/reg&       0.06    &   0.04\\
  17  58  58.78  &  -29  10  49.51  &    22.60  &    20.94    &  var/reg&       2.27    &   0.07\\
  17  58  59.17  &  -29  10  21.82  &    21.64  &    20.19    &  var/reg&       1.26    &   0.09\\
  17  58  55.42  &  -29  13   9.91  &    24.34  &    22.33    &  var/reg&       1.78    &   0.35\\
  17  58  57.06  &  -29  13  35.16  &    20.58  &    19.35    &var/irreg&       \nodata    &   0.09\\
  17  58  57.21  &  -29  11  18.19  &    21.32  &    19.94    &var/irreg&       \nodata    &   0.04\\
  17  58  59.34  &  -29  10  35.30  &    21.78  &    20.31    &  var/reg&       0.38    &   0.07\\
  17  58  59.68  &  -29  13  14.44  &    22.33  &    20.65    &  var/reg&       0.94    &   0.14\\
  17  58  53.72  &  -29  13  28.78  &    24.02  &    22.22    &  var/reg&       0.65    &   0.25\\
  17  58  54.91  &  -29  10  30.76  &    20.01  &    18.73    &var/irreg&       \nodata    &   0.06\\
  17  58  55.71  &  -29  13  30.64  &    20.90  &    19.66    &  var/reg&       2.44    &   0.09\\
  17  58  54.44  &  -29  10  35.83  &    22.37  &    20.88    &  var/reg&       4.81    &   0.10\\
  17  58  55.79  &  -29  10  57.05  &    23.00  &    20.64    &  var/reg&       2.84    &   0.19\\
  17  58  54.17  &  -29  13  21.45  &    21.23  &    19.99    &  var/reg&       2.09    &   0.05\\
  17  58  54.90  &  -29  12  27.10  &    20.27  &    19.14    &var/irreg&       \nodata    &   0.03\\
  17  59   0.19  &  -29  11  32.85  &    23.06  &    20.34    &  var/reg&       0.83    &   0.11\\
  17  58  55.06  &  -29  13  12.33  &    23.95  &    21.98    &  var/reg&       2.01    &   0.12\\
  17  58  59.71  &  -29  12  17.19  &    22.55  &    21.05    &  var/reg&       1.04    &   0.11\\
  17  58  54.44  &  -29  11  45.74  &    22.34  &    20.93    &  var/reg&       7.60    &   0.08\\
  17  58  59.53  &  -29  10  29.80  &    20.10  &    18.75    &var/irreg&       \nodata    &   0.09\\
  17  58  57.20  &  -29  11  57.55  &    23.01  &    21.37    &  var/reg&       0.38    &   0.16\\
  17  58  56.44  &  -29  11  13.40  &    23.25  &    21.51    &  var/reg&       6.27    &   0.15\\
  17  58  55.67  &  -29  10  56.72  &    25.68  &    24.40    &  var/reg&       2.85    &   1.27\\
  17  58  52.67  &  -29  13  32.92  &    21.63  &    19.85    &var/irreg&       \nodata    &   0.12\\
  17  59   2.88  &  -29  11   2.27  &    20.77  &    19.53    &  var/reg&       1.60    &   0.09\\
  17  59   1.11  &  -29  12  47.56  &    21.46  &    20.07    &var/irreg&       \nodata    &   0.07\\
  17  59   3.63  &  -29  13  19.71  &    22.35  &    19.96    &var/irreg&       \nodata    &   0.06\\
  17  59   7.09  &  -29  12  26.42  &    21.27  &    18.73    &  var/reg&       1.10    &   0.04\\
  17  59   5.64  &  -29  13  10.35  &    21.83  &    20.50    &  var/reg&       1.98    &   0.09\\
  17  59   1.13  &  -29  10  27.93  &    22.14  &    20.68    &  var/reg&       0.90    &   0.07\\
  17  59   7.54  &  -29  12  59.05  &    21.08  &    18.79    &var/irreg&       \nodata    &   0.05\\
  17  59   5.75  &  -29  10  57.46  &    20.01  &    18.87    &var/irreg&       \nodata    &   0.03\\
  17  59   3.66  &  -29  13  30.80  &    20.84  &    19.62    &  var/reg&       5.95    &   0.05\\
  17  59   1.33  &  -29  10  48.18  &    21.62  &    20.19    &  var/reg&       1.29    &   0.10\\
  17  59   7.17  &  -29  10  21.71  &    22.26  &    20.74    &var/irreg&       \nodata    &   0.05\\
  17  59   4.85  &  -29  11  15.24  &    22.05  &    20.33    &  var/reg&       1.35    &   0.18\\
  17  59   5.54  &  -29  10  44.69  &    21.44  &    20.14    &  var/reg&       0.90    &   0.12\\
  17  59   7.47  &  -29  13  22.25  &    22.22  &    19.97    &  var/reg&       6.41    &   0.07\\
  17  59   4.36  &  -29  13   7.39  &    21.18  &    19.90    &  var/reg&       0.65    &   0.08\\
  17  59   3.71  &  -29  11  56.65  &    20.16  &    19.05    &  var/reg&       0.31    &   0.02\\
  17  59   2.21  &  -29  10  51.44  &    20.98  &    19.67    &var/irreg&       \nodata    &   0.05\\
  17  59   3.80  &  -29  12  41.39  &    21.46  &    20.29    &var/irreg&       \nodata    &   0.05\\
  17  59   1.80  &  -29  13  24.95  &    21.48  &    20.28    &  var/reg&       0.52    &   0.04\\
  17  59   7.31  &  -29  10  55.30  &    20.08  &    18.96    &  var/reg&       6.00    &   0.02\\
  17  59   3.97  &  -29  13  11.61  &    23.14  &    21.50    &  var/reg&       1.19    &   0.31\\
  17  59   7.69  &  -29  12  38.15  &    23.41  &    21.70    &  var/reg&       0.87    &   0.17\\
  17  59   5.87  &  -29  13  18.64  &    22.24  &    20.56    &  var/reg&       0.60    &   0.13\\
  17  59   7.60  &  -29  13   8.61  &    20.14  &    19.01    &var/irreg&       \nodata    &   0.03\\
  17  59   7.99  &  -29  11  12.31  &    21.78  &    20.27    &  var/reg&       1.22    &   0.10\\
  17  59   8.41  &  -29  10  20.53  &    22.53  &    21.08    &  var/reg&       1.66    &   0.14\\
  17  59   4.22  &  -29  11   8.95  &    22.05  &    20.38    &var/irreg&       \nodata    &   0.11\\
  17  59   8.49  &  -29  12  36.43  &    22.83  &    21.23    &  var/reg&       2.78    &   0.13\\
  17  59   3.37  &  -29  11  29.90  &    24.22  &    22.30    &  var/reg&       0.71    &   0.19\\
  17  59   4.86  &  -29  12   4.81  &    21.05  &    19.74    &  var/reg&       1.18    &   0.06\\
  17  59   7.02  &  -29  10  55.17  &    22.19  &    20.62    &  var/reg&       0.61    &   0.15\\
  17  59   5.47  &  -29  12   9.13  &    20.75  &    19.47    &  var/reg&       0.85    &   0.08\\
  17  59   2.63  &  -29  13  34.60  &    22.52  &    20.99    &var/irreg&       \nodata    &   0.09\\
  17  59   3.58  &  -29  10  59.49  &    21.77  &    20.30    &  var/reg&       2.02    &   0.07\\
  17  59   0.82  &  -29  12  30.11  &    22.41  &    20.92    &  var/reg&       1.87    &   0.11\\
  17  59   1.38  &  -29  12  58.27  &    22.91  &    21.32    &var/irreg&       \nodata    &   0.20\\
  17  59   6.17  &  -29  11  59.10  &    22.72  &    21.02    &  var/reg&       2.84    &   0.10\\
  17  59   5.17  &  -29  10  20.90  &    21.87  &    20.47    &  var/reg&       0.84    &   0.10\\
  17  59   4.23  &  -29  12  52.31  &    23.32  &    20.48    &  var/reg&       0.66    &   0.13\\
  17  59   7.00  &  -29  13   1.54  &    25.32  &    24.26    &  var/reg&      11.79    &   0.77\\
  17  59   4.73  &  -29  11  58.55  &    24.51  &    22.38    &  var/reg&       1.54    &   0.29\\
  17  59   4.47  &  -29  11  30.23  &    22.89  &    21.28    &var/irreg&       \nodata    &   0.12\\
\enddata
\tablenotetext{a}{Range is defined as the difference in amplitude between the values in the 95th and 5th percentile
in the original light curve; see \S 4.2 for details.}
\end{deluxetable}

\end{document}